\documentclass[final,1p,times,twocolumn]{elsarticle}
\usepackage{epsfig}
\usepackage{amssymb}
\usepackage{amsmath}
\usepackage{mathtools}
\usepackage{longtable}
\usepackage{subcaption}
\journal{Nuclear Physics B}
\usepackage{xcolor}

\DeclareMathAlphabet{\oldcal}{OMS}{cmsy}{m}{n}
\newcommand{\bigo}[1]{\oldcal{O}\left(#1\right)}
\usepackage{hyperref}

\begin{document}
\title{Universal effective couplings of the three-dimensional
$n$-vector model and field theory}

\author[label1]{\corref{cor2}A.\,Kudlis}
\ead{andrewkudlis@gmail.com}\cortext[cor2]{Corresponding author}

\author[label1]{A.\,I.\,Sokolov}

\address[label1]{St. Petersburg State University, 7/9 Universitetskaya Emb., St. Petersburg, 199034 Russia}

\date{\today}

\begin{abstract}
We calculate the universal ratios $R_{2k}$ of renormalized coupling
constants $g_{2k}$ entering the critical equation of state for the
generalized Heisenberg (three-dimensional $n$-vector) model.
Renormalization group (RG) expansions of $R_8$ and $R_{10}$ for arbitrary
$n$ are found in the four-loop and three-loop approximations respectively.
Universal octic coupling $R_8^*$ is estimated for physical values of spin
dimensionality $n = 0, 1, 2, 3$ and for $n = 4,...64$ to get an idea about
asymptotic behavior of $R_8^*$. Its numerical values are obtained by means
of the resummation of the RG series and within the pseudo-$\varepsilon$
expansion approach. Regarding $R_{10}$ our calculations show that three-loop 
RG and pseudo-$\varepsilon$ expansions possess big and rapidly growing 
coefficients for physical values of $n$ what prevents getting fair 
numerical estimates. 

\end{abstract}

\begin{keyword}
renormalization group, multi-loop calculations, effective coupling
constants, universal ratios.

\MSC{82B28}
\end{keyword}

\maketitle

\section{Introduction}

The thermodynamics of O($n$)-symmetric systems near the Curie point is
characterized by universal parameters that depend only on the ordering
field dimensionality $n$ and the number of space dimensions D. The
renormalized effective coupling constants $g_{2k}$ and the ratios $R_{2k}
= g_{2k}/g_4^{k-1}$ entering the small-magnetization expansion of the free
energy (the effective potential)
\begin{equation}
\label{eq1}
F(z,m) - F(0,m) = {\frac{m^3 }{g_4}} \Biggl({\frac{z^2 }{2}} + z^4 + R_6
z^6 + R_8 z^8 + R_{10} z^{10}... \Biggr),
\end{equation}
are among them. They determine, along with effective potential, the
nonlinear susceptibilities $\chi_{2k}={\partial^{2k-1}
M_{\alpha}}/{\partial H_{\alpha}^{2k-1}}$ above $T_c$:
\begin{eqnarray}
\label{eq2} \chi_4 = - 24{\frac{\chi^2}{m^3}}g_4, \qquad \chi_6 = -
6!{\frac{\chi^3 g_4^2}{m^6}}(R_6-8), \qquad
\chi_8 = - 8!{\frac{\chi^4 g_4^3}{m^9}}(R_8-24 R_6+96), \\
\nonumber \chi_{10} = -10!{\frac{\chi^5 g_4^4}{m^{12}}}(R_{10}-32 R_8-18
R_6^2+528 R_6-1408).
\end{eqnarray}
Here $z = M \sqrt{g_4/m^{1 + \eta}}$ is a dimensionless magnetization, $m
\sim (T - T_c)^\nu$ being an inverse correlation length, $\chi$ is a
linear susceptibility and $\chi_4$, $\chi_6$, $\chi_8$, $\chi_{10}$ are
nonlinear susceptibilities of corresponding orders.

For the three-dimensional (3D) $n$-vector and related models, the
effective potential and nonlinear susceptibilities are intensively studied
during several decades. Renormalized coupling constants $g_{2k}$ and
universal ratios $R_{2k}$ were evaluated by a variety of analytical and
numerical methods
\cite{PhysRevB.15.1552,PhysRevLett.39.95,PhysRevB.17.1365,PhysRevB.21.3976,PhysRevB.41.402,PhysRevLett.68.3674,PhysRevD.48.4919,TETRADIS1994541,PhysRevD.51.1875,REISZ199577,PhysRevE.51.1894,CAMPOSTRINI1996207,S96,ZLF,SOU,GZ97,Morr,BC97,GZJ98,SO98,BC98,PV98,PV,S98,SOUK99,CPRV99,PV2000,PS2000,CPV2000,PS2001,ZJ01,CHPV2001,CHPRV2001,CHPRV2002,CPRV2002,ZJ,PV02,TPV03,BT2005,CSS07,BP11,NS14,KKN16,KS16,K16,KS17,AKL17,SNK17,KP17,LK17}.
Estimating universal critical values of quartic and sextic effective
couplings by means of the field-theoretical renormalization group (RG)
approach has shown that RG technique enables one to get rather accurate
numerical results for these quantities. Multi-loop RG calculations of
$g_4$ and $g_6$ ($R_6$) combined with proper resummation procedures were
found to yield numerical estimates that have three-four decimals level of
accuracy. For example, advanced RG estimates of the Wilson fixed point
location $g = g_4(n+8)/2\pi$ for $n = 1$ lie between 1.411 and 1.417
\cite{PhysRevLett.39.95,PhysRevB.17.1365,PhysRevB.21.3976,GZJ98} while field-theoretical results $R_6 =
1.648$ and $R_6 = 1.649$ \cite{GZ97,SOUK99,SNK17} perfectly agree with the
value $1.649 \pm 0.002$ given by advanced lattice calculations
\cite{BP11}.

The attempts to get accurate enough field-theoretical estimates for
higher-order coupling $g_8$ and universal ratio $R_8$ were, however, much
less successful. The numbers obtained by means of resummation of the
three-loop RG series for $g_8$ \cite{SOUK99} and extracted from
corresponding pseudo-$\varepsilon$ expansions for $g_8$ and $R_8$ turn out to
differ markedly from each other and from known lattice estimates (see, e.
g. Table X in Ref. \cite{NS14} where the relevant data are collected). In
principle, this is not surprising since the RG expansion for octic
coupling found in Ref. \cite{SOUK99} is shorter than those for $g_6$
(four-loop) and for $\beta$ function (six-loop) resulting in numerical
values of $g_4$. Moreover, it is stronger divergent. For $g_8$, however,
there is an extra reason -- the last but not least -- making existing
numerical estimates unexpectedly crude. The point is that the RG series
for $g_8$ has an unusual feature: the first two terms in this series tend
to compensate each other both for physical values of $n$ and for $n \gg 1$
\cite{SOUK99}. This makes their mutual contribution small and increases
the role of the higher-order terms. The same is true for corresponding
pseudo-$\varepsilon$ expansions \cite{NS14}. So, finding of higher-order
terms in the RG series appears to be extremely important for getting
proper numerical estimates of the octic coupling.

In this paper, we extend the RG expansion of effective coupling constants
$g_8$ up to four-loop order and calculate three-loop RG series for
$g_{10}$ under arbitrary $n$. Since the pseudo-$\varepsilon$ expansion
approach was shown to be highly efficient when used to evaluate critical
exponents and other universal quantities \cite{PhysRevB.21.3976,NS14,KS16,KS17,FH97,FH99,FHY2000,HDY01,DHY02,CPN04,CP04,COPS04,HID04,DHY04,CP05,S2005,S2013,NS13,NS14e,NS16,NS16h} we calculate the pseudo-$\varepsilon$ expansions for
both couplings as well. To get reliable numerical estimates for higher-order 
couplings and universal ratios the Wilson fixed point location is refined 
using the resummation procedures that allow optimization by adjusting 
free parameters involved. The RG series and corresponding pseudo-$\varepsilon$ 
expansion for the octic coupling are then resummed by means of 
Borel-transformation-based techniques, and the numbers obtained are compared
with their counterparts found within the alternative approaches. The structure 
of RG series and pseudo-$\varepsilon$ expansions for $g_{10}$ and $R_{10}$ is 
discussed and the conclusion concerning the capability of the field theory 
to give proper numerical estimates for this coupling is made.

\section{Perturbative, RG and pseudo-\texorpdfstring{$\varepsilon$}{Lg} expansions}

The critical behavior of 3D $n$-vector model is described by Euclidean
field theory with the Hamiltonian:
\begin{equation}
H = \int d^{3}x \Biggl[\frac{1}{2}\left( m_0^2 \varphi_{\alpha}^2 +
(\nabla \varphi_{\alpha}^2)\right) + \frac{\lambda}{24}
(\varphi_{\alpha}^2)^2 \Biggr],
\end{equation}
where $\varphi_{\alpha}$ is a real vector field, $\alpha = 1,...n$, bare
mass squared $m_0^2$ being proportional to $T - T_c^{(0)}$ and $T_c^{(0)}$
-- mean field transition temperature.

To derive RG expansions for $g_8$ and $g_{10}$ we employ the following --
straightforward -- method. The renormalized perturbative series are found
within the massive theory from conventional Feynman graph expansions for
eight-point and ten-point vertices in terms of the quartic coupling
constant $\lambda$. In the course of such calculations the tensor
structure of these vertices is taken into account:
\begin{equation}
\Gamma _{\alpha \beta \gamma \delta \mu \nu \rho \sigma} ={\frac 1{105}}
(\delta _{\alpha \beta }\delta _{\gamma \delta }\delta _{\mu \nu }\delta
_{\rho \sigma }+104~permutations)\Gamma_8,
\label{eq:4} \\
\end{equation}
\begin{equation}
\Gamma _{\alpha \beta \gamma \delta \mu \nu \rho \sigma \xi \zeta }={\frac
1{945}}(\delta _{\alpha \beta }\delta _{\gamma \delta }\delta _{\mu \nu}
\delta _{\rho \sigma } \delta _{\xi \zeta }+
944~permutations)\Gamma_{10},
\label{eq:5} \\
\end{equation}
where $\Gamma_{2k} = g_{2k} m^{3 - k(1 + \eta)}$, $m$ and $\eta$ are
renormalized mass and Fisher exponent respectively. The bare coupling
constant $\lambda$, in its turn, is expressed perturbatively as a function
of the renormalized dimensionless coupling $g_4$. Substituting
corresponding power series for $\lambda $ into original expansions we
obtain the RG series for $g_8$ and $g_{10}$.

The one-, two-, three- and four-loop contributions to $g_8$ are formed by
1, 5, 36 and 268 one-particle irreducible Feynman graphs, respectively. 
Corresponding integrals, symmetry and tensor factors are presented 
in Table 1 of Supplementary materials (see~\ref{app:suppl}) where Nickel's notations describing graphs topology are used. Summing up all the calculated graphs we obtain:
\begin{equation}
\begin{split}
g_8=&-\frac{81}{2\pi}\left(\frac{\lambda
Z^2}{m}\right)^4\left[\frac{n+80}{81} -\frac{405 n^2+ 35626 n +
342320}{13122\pi}\left(\frac{\lambda Z^2}{m}\right)
+ \left(0.0046907955 n^3 \right. \right.\\
&+ \left. 0.463650683 n^2 + 8.86811653 n + 45.4769028 \right)
\left(\frac{\lambda Z^2}{m}\right)^2 + \left(0.00174198 n^4\right.\\
&\left.\left. + 0.194893055 n^3 + 5.58218793 n^2 + 59.25883462 n +
209.3927445\right)\left(\frac{\lambda Z^2}{m}\right)^3\right].
\end{split}
\end{equation}
The expansion for $\lambda$ in terms of renormalized dimensionless
effective coupling $g_4$ emerges directly from the normalizing condition
$\lambda = m Z_4 Z^{-2} g_4$ and the known series for $Z_4$ \cite{PhysRevE.51.1894}:
\begin{eqnarray}
Z_4 &=&1+{\frac{n+8}{{2\pi }}}g_4+{\frac{3~n^2+38~n+148}{{12\pi ^2}}}%
g_4^2\qquad \qquad \qquad \qquad \qquad \qquad \qquad \qquad \qquad
\nonumber \\
&+&(0.0040314418~n^3+0.0679416657~n^2+0.466356233~n+1.240338484)g_4^3.
\label{eq:7}
\end{eqnarray}
Combining these expressions we obtain
\begin{equation}
\begin{split}
g_8 = &-\frac{81}{2\pi} g_4^4 \left[ \frac{n+80}{81}-\frac{81 n^2+7114 n
+134960}{13122 \pi}g_4 + \left(0.00943497 n^2 \right.\right.\\
&+\left. 0.60941312 n + 7.15615323 \right)g_4^2
- \left(0.00013078 n^3+0.04703841 n^2 \right. \\
&+\left.\left. 1.97176517 n+16.56483375\right)g_4^3\right].
\end{split}
\end{equation}
The coefficients of the series in square brackets are seen to grow
progressively with $n$. On the other hand, the Wilson fixed point
coordinate $g_4^*$ is known to diminish as a function of this variable.
That is why in order to make the numerical structure of RG expansions for
higher-order couplings more transparent it is natural to use, instead of
$g_4$, the rescaled quartic coupling constant
\begin{equation}
g = {\frac{{n + 8}}{{2\pi}}}g_4,
\label{eq:9}
\end{equation}
whose critical (fixed point) value only weakly depends on $n$ lying
between 1.42 and 1 for any $n$. The RG series for $g_8$ in terms of $g$ is
as follows:
\begin{equation}
\begin{split}
g_8 = &-\frac{648\pi^3}{(n+8)^4} g^4 \left[ \frac{n+80}{81}
-\frac{g}{(n+8)}\frac{81 n^2+7114 n+134960}{6561}
+ \frac{g^2}{(n+8)^2}\left(0.372477560 n^2 \right.\right.\\
&+\left. 24.0586651 n + 282.513606\ \right)
- \frac{g^3}{(n+8)^3}\left(0.0324416663 n^3+ 11.6678970 n^2 \right. \\
&+\left.\left. 489.096788 n + 4108.91056\right)\right].
\end{split}
\end{equation}

In the case of $g_{10}$, the one-, two-, and three-loop contributions are
given by 1, 6 and 64 Feynman graphs, respectively. They are listed in 
Table 2 of Supplementary materials (see~\ref{app:suppl}). Corresponding "bare" and renormalized 
perturbative expansions are found to be:
\begin{equation}
\begin{split}
g_{10}=&\frac{243}{\pi}\left(\frac{\lambda
Z^2}{m}\right)^5\left[\frac{n+242}{243} -\frac{567 n^2 + 128108 n +
1544380}{39366\pi}\left(\frac{\lambda Z^2}{m}\right)
+ \left(0.0029187172 n^3 \right. \right.\\
&+\left. \left. 0.64392821 n^2 + 15.53330379 n + 94.77350036 \right)
\left(\frac{\lambda Z^2}{m}\right)^2\right],
\end{split}
\end{equation}

\begin{equation}
\begin{split}
g_{10} =&\frac{243}{\pi} g_4^5 \left[ \frac{n + 242}{243}
- \frac{81 n^2 + 13429 n +380150}{19683\pi} g_4 + \right. \\
&+\left. \left(0.0001042399 n^3 +0.02139208 n^2 +1.42119791 n +
21.74148152 \right)g_4^2 \right].
\end{split}
\end{equation}
The latter series in terms of $g$ reads:
\begin{equation}
\begin{split}
g_{10} =&\frac{7776\pi^4}{(n+8)^5}g^5\left[ \frac{n + 242}{243}
- \frac{162 n^2 + 26858 n + 760300}{19683(n+8)} g + \right. \\
&+\left. \left(0.00411522634 n^3 +0.844525548 n^2 + 56.1066447 n +
858.319287 \right)\frac{g^2}{(n+8)^2} \right].
\end{split}
\end{equation}

The RG series derived turn out to be strongly divergent. To make them more
suitable for getting numerical estimates the pseudo-$\varepsilon$ expansion
technique may be employed. Pseudo-$\varepsilon$ expansions for the critical
values of $g_8$ and $g_{10}$ can be derived from (8) and (12) substituting
the pseudo-$\varepsilon$ expansion for the Wilson fixed point coordinate
which is as follows \cite{NS14}:
\begin{eqnarray}
\label{g4-tau}
g_4^* &=& \frac{2\pi}{n+8}\biggl[\tau + \frac{\tau^2}{(n +
8)^2} \biggl(6.074074074~n + 28.14814815 \biggr)
\nonumber \\
&+& \frac{\tau^3}{(n + 8)^4} \biggl(- 1.34894276~n^3 + 8.056832799~n^2 +
44.73231547~n - 12.48684745 \biggr)
\nonumber \\
&+& \frac{\tau^4}{(n + 8)^6} \biggl( - 0.15564589~n^5 - 7.638021730~n^4 +
100.0250844~n^3 + 679.8756744~n^2
\nonumber \\
&+& 1604.099837~n + 3992.366079 \biggr) \biggr].
\end{eqnarray}
Combining (8), (12) and (14) we obtain:
\begin{equation}
\begin{split}
g_8^* =&-\frac{8\pi^3}{(n+8)^4}\tau^4\left[n + 80 + \frac{\tau}{(n+8)^2}\left( 248.05021 n^3 + 17743.246 n^2
+ 77514.161 n + 1072066.9\right) \right. \\
&+\frac{\tau^2}{(n+8)^4}\left(1387.9548 n^4 + 197852.87 n^3 + 1715306.9
n^2
+ 15922971 n + 8711449.0\right) \\
&+ \frac{\tau^3}{(n+8)^6}\left(- 866.7700 n^6 - 37159.01 n^5 +
1.204930\cdot10^6 n^4
+ 2.880413\cdot10^7 n^3  \right. \\
&+\left. 4.461346\cdot10^8 n^2 + 2.163345\cdot10^9 n +
4.901591\cdot10^9\right) \left.\right],
\end{split}
\end{equation}
\begin{equation}
\begin{split}
g_{10}^* =&\frac{32\pi^4}{(8 + n)^5}\tau^5\left[n + 242 - \frac{\tau}{(n+8)^2}\left(6234.18183 n^3+ 988772.024 n^2+ 14178684.5 n
+127900782 \right) \right. \\
&+\frac{\tau^2}{(n+8)^4}\left(\right. 3117.09091 n^5  + 441336.763 n^4
+ 8582256.44 n^3 +140472393 n^2\\
&+ 1.02642700\cdot10^9 n + 8.00714134\cdot10^9 \left.\right) \left.\right]. \\
\end{split}
\end{equation}
Since the equation of state and expressions for nonlinear susceptibilities
contain the universal ratios rather than effective coupling constants
themselves it is reasonable to have the pseudo-$\varepsilon$ expansions for
the critical values of these ratios. They are as follows:
\begin{eqnarray}
R_8^*&=&-\frac{n+80}{n+8}\tau +{\frac{\tau^2}{(n+8)^3}}(n^3 +
89.753086
n^2 + 1854.7160 n + 11077.531) \\
\nonumber &-&{\frac{\tau^3}{(n+8)^5}}(16.67359 n^4+ 1111.2054 n^3 +
22512.707 n^2
+ 199142.42 n + 713156.70) \\
\nonumber &+&{\frac{\tau^4}{(n+8)^7}}(0.0855557 n^6 + 272.9566 n^5 +
15580.31 n^4
+363790.3 n^3 \\
\nonumber &+&4151127 n^2 + 2.348384\cdot10^7 n + 5.664042\cdot10^7),
\end{eqnarray}
\begin{eqnarray}
R_{10}^*&=&\frac{2(n+242)}{n+8}\tau - \frac{\tau^2}{(n+8)^3}(4 n^3
+ 683.012346 n^2 + 21081.9753 n + 136559.012) \\
\nonumber &+&\frac{\tau^3}{(n+8)^5}(2 n^5 + 391.148939 n^4 + 24655.9875
n^3
+ 552045.964 n^2 \\
\nonumber &+&5261106.26 n + 18236388.8).
\end{eqnarray}
These RG expansions and $\tau$-series will be used to estimate
higher-order effective couplings near the Curie point.

\section{Wilson fixed point coordinate}

Before going further to the numerical analysis of the universal ratios we would 
like to perform some refinement of the Wilson fixed point coordinate for different $n$ 
which was calculated earlier in a number of papers (see, e.g. 
\cite{PhysRevLett.39.95,PhysRevB.17.1365,PhysRevB.21.3976,PhysRevE.51.1894,GZJ98,S98,SOUK99,NS14}). For this purpose the authors 
developed certain procedures based on both Pad\'e--Borel--Leroy (PBL) and conform-Borel 
(CB) resummation techniques. Here we briefly describe the steps of the algorithms with 
the specific attention paid to the choice of the fitting parameters involved.

Let us start with the first -- PBL -- technique which is a transparent and concise procedure. 
As a starting point we take the $\beta$ function of the $O(n)$-symmetric model obtained 
within the RG perturbation theory in three dimensions. Knowing the asymptotic behaviour 
of the expansion coefficients for $\beta$ function we can apply the Borel 
transformation which factorially diminishes these coefficients making the transformed 
series —- the Borel image -— convergent with finite radius of convergence. 
This expansion when being complete, i. e. having infinite number of perturbative terms, 
would converge to some analytical function which may be considered as an analytical  
continuation of the series sum beyond the convergence radius. In practice, however, 
only truncated series are in hand, so we have to perform the analytical continuation 
approximately, in some reasonable way using all the information available. Within the PBL resummation approach Pad\'e approximants being an example of classical method 
of rational approximation are employed for the analytical continuation. Schematically, 
this procedure looks as follows:
\begin{eqnarray} \label{pbl_pf_proc}
\label{pbl_proc}
    \beta_{b,res}^{N,[L/M]}(g)=\int\limits_{0}^{\infty}dt e^{-t}t^b[L/M][B^{N}_{b}[\beta]](gt), \  B^{N}_{b}[\beta](g)=\sum\limits_{k=0}^{N}\frac{\beta_k}{\Gamma(k+b+1)}g^k, \  \beta(g)=\sum\limits_{k=0}^{\infty}\beta_kg^k, \\ \nonumber
    [L/M][B^{N}_{b}[\beta]](g)=\frac{a_0(\{\beta_i\}_{i}^N) + a_1(\{\beta_i\}_{i}^N)g+a_2(\{\beta_i\}_{i}^N)g^2+ \dots + a_L(\{\beta_i\}_{i}^N)g^L}{1+b_1(\{\beta_i\}_{i}^N)g+b_2(\{\beta_i\}_{i}^N)g^2+ \dots + b_M(\{\beta_i\}_{i}^N)g^M}, \ \ L+M=N.
\end{eqnarray}
In order to evaluate the Wilson fixed point coordinate $g^*$ in the six-loop (highest-order 
available) approximation, we have to find the non-trivial root of the following equation:
\begin{equation}
    \beta(g^*)=0, \ \ \beta_{b,res}^{6,[L/M]}(g(b,L,M))=0, \ \ g^*=g(b^*,M^*,N^*).
\end{equation}
Obviously, with this set of parameters -- $\{L,M,b\}$ -- we have some ambiguity 
in the course of finding $g^*$. How to choose properly the values of these parameters? As for 
fitting parameter $b$, it accelerates the convergence of the estimates. However, since 
it has no physical meaning, when choosing its specific numerical value it is natural 
to require the least sensitivity of the function $g(b,L,M)$ in the vicinity of $b^*$ 
with respect to $L$ and $M$. As is well known the diagonal -- $[L/L]$ --  and 
near-diagonal -- $[L/L-1], [L-1/L]$ -- approximants possess the best approximating 
properties. However, there are no obvious preferences between them except for the fact 
that the approximant should have no poles on positive real axis which would cause 
divergences in integration (\ref{pbl_pf_proc}). That is why here we require also 
a minimal variation of the estimate of the fixed point location when using certain 
approximant. So, our way of doing is as follows. 

\begin{figure}[h!]
\centering
\includegraphics[scale=0.82]{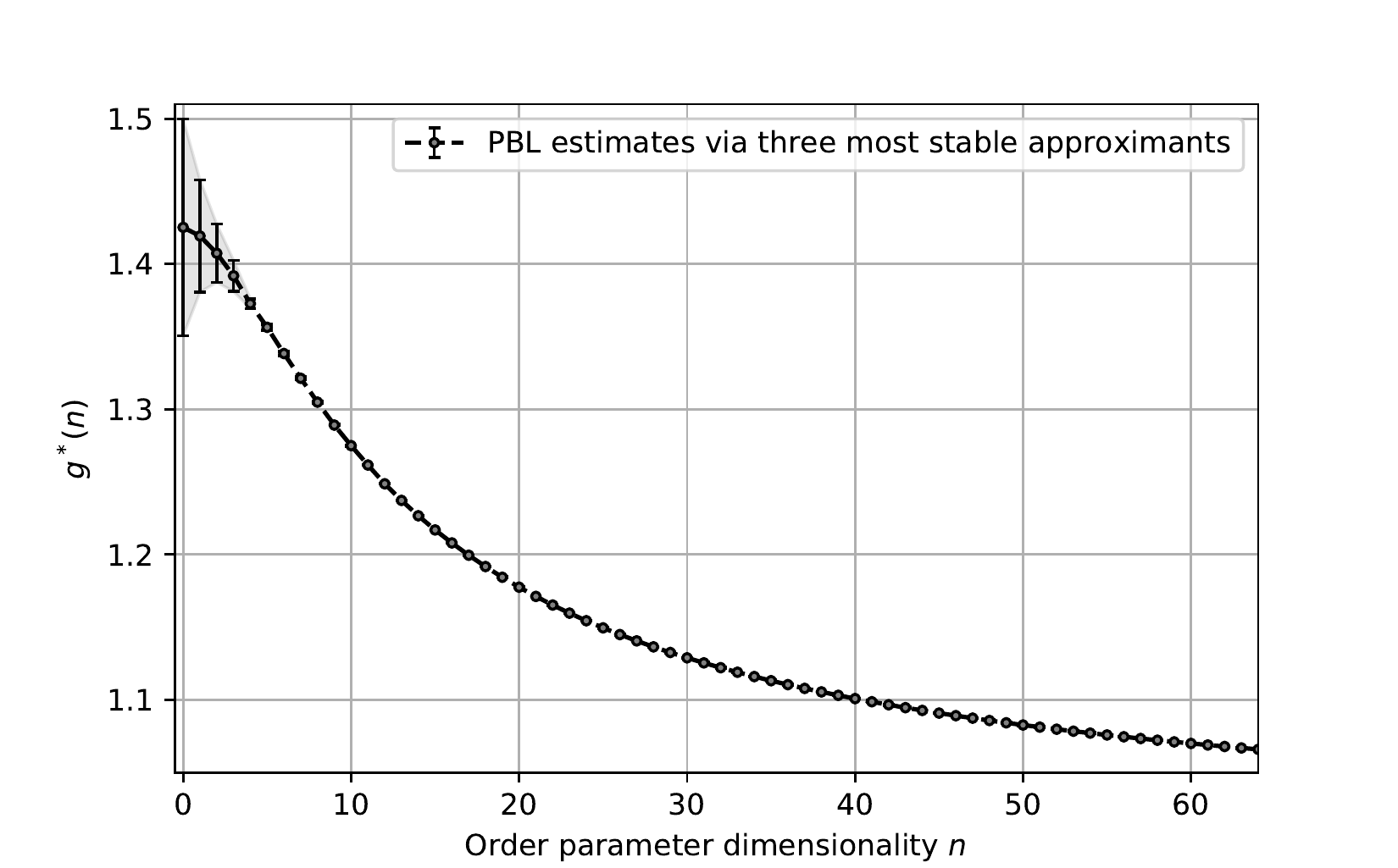}
\caption{PBL estimates of the Wilson fixed point coordinate $g^*$ for various $n$.}
\label{fp_pbl_0_64}
\end{figure}

We define reasonable interval of variation of parameter $b$, choose a step with which 
we will analyze it and calculate a \textit{Pad\'e triangle} for each point $b$, the example of which is presented in Table \ref{fp_pbl_triangle_for_n=6}.
\begin{table}[h!]
\centering
\caption{Example of Pad\'e triangle with PBL estimates of fixed-point coordinate for $n=6$. The optimal value of shift parameter $b$ is $0.01$. '-' indicates that corresponding approximant is not relevant.}
\label{fp_pbl_triangle_for_n=6}
\renewcommand{\tabcolsep}{0.3cm}
\begin{tabular}{{c}|*{7}{c}}
$M \setminus L$ & 0 & 1 & 2 & 3 & 4&5&6 \\
\hline
0&  -  & 1      &  -	& 1.118 &    -   &  1.098 & -    \\ 
1&  -  & 1.426  & 1.331    & 1.338 &  1.340 &  1.337 &   \\
2&  -  & 1.294  & 1.338	& 1.341 &  1.339 & 	  &   \\
3&  -  & 1.359  & 1.340 	& 1.339 & 	 & 	  &   \\
4&  -  & 1.348  & 1.336 	& 	& 	 & 	  &   \\
5&  -  & 1.339  & 		& 	& 	 & 	  &   \\
6&  -  & 	     & 		&	& 	 & 	  &   \\
\hline
\end{tabular}
\end{table}
With these triangles in hand we construct some sample which is formed by the most reliable 
approximants. What we have to include into this set? In addition to the highest-order 
approximants, in order to somehow take into account fluctuations of convergence which 
are believed to decrease with increasing order, we have to include into consideration 
the lower-order ones. Then, as was already said, we exclude approximants spoiled by the 
positive axis poles. Since boundary approximants $[L/0]$ and 
$[0/L]$ usually contribute big errors they are also excluded. The results of such 
calculations for various $n$ are presented in Table \ref{tab1} and in Figure 
\ref{fp_pbl_0_64}. 

\begin{figure}[!ht]
\centering
\includegraphics[scale=0.72]{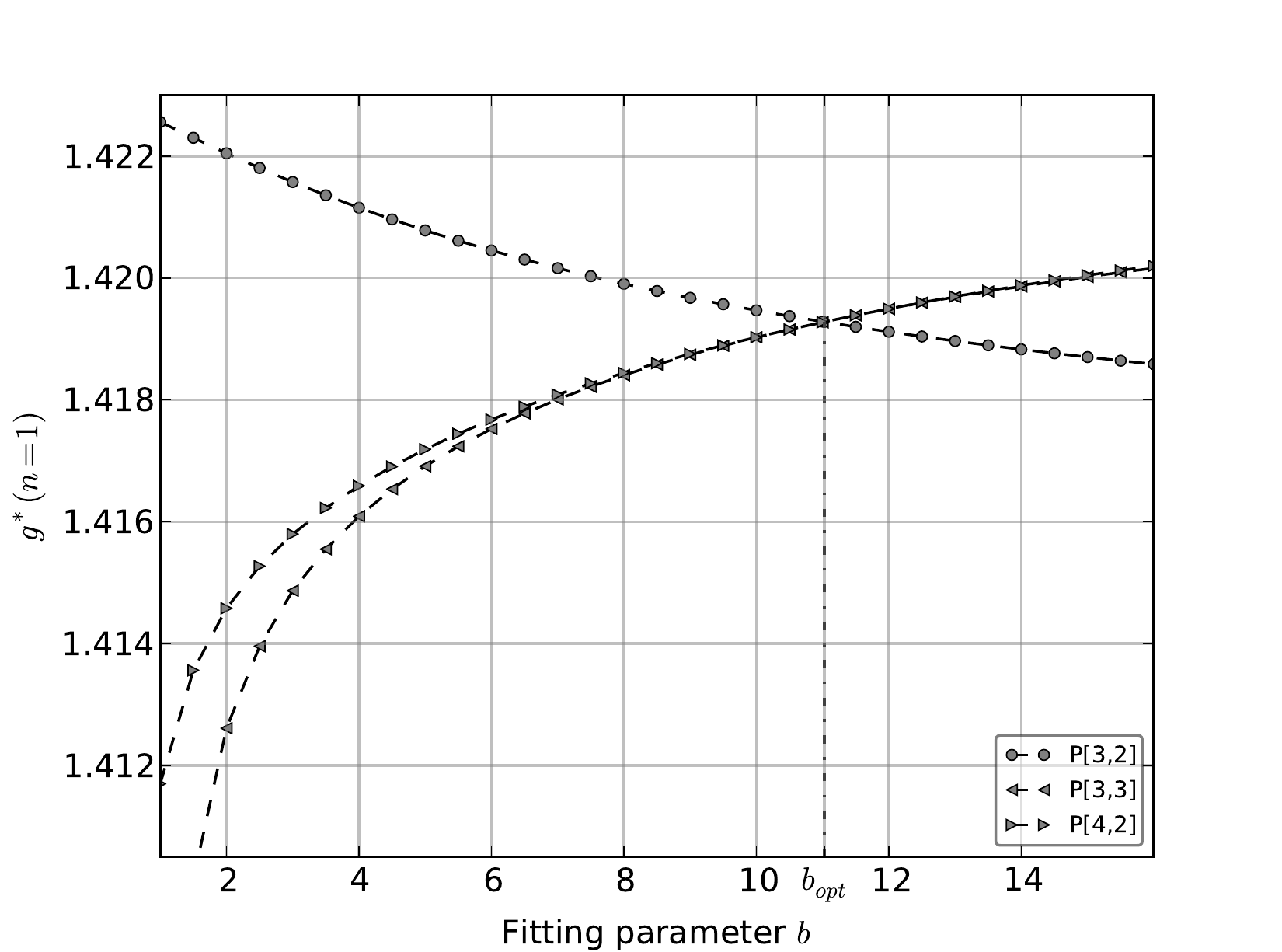}
\caption{The values of $g^*$ given by three most stable PBL 
approximants under $n=1$ as functions of $b$. The intersection 
yields $g^*=1.4194$.}
\label{PBL_conv_fp_n=1_3best}
\end{figure}

Also, for clarity, we demonstrate in Figure \ref{PBL_conv_fp_n=1_3best} the  
dependence of $g^*$ given by three most stable PBL approximants on fitting parameter $b$.

If we have the information about the asymptotic behavior of the series coefficients 
we can resort to the CB resummation technique. For $\varphi^4$ theory, the Lipatov's asymptotics is well known~\cite{Lipatov1977,PhysRevD.15.1544}:
\begin{equation}\label{lip_as}
    \beta_k \propto a^k \ k! \ k^{b_{\beta}}, \ \ k\rightarrow \infty.
\end{equation}
It allows to construct an analytical continuation of the Borel image that is expected to have efficient approximating properties. 
The detailed description of CB resummation procedure was given in many works (see, e.g., \cite{PhysRevB.21.3976,GZ97,GZJ98,ZJ01,ZJ,PV02}). Here we restrict ourselves with the working 
formulas and description of the algorithm of choosing the resummation parameters. 
The machinery used by authors is the following. First, we separate in series for 
$\beta$ function its "loop" (perturbative) part. Within the normalization we use, 
in the critical point this part has to coincide with negative value of the Wilson 
fixed point coordinate ($-g^*$):
\begin{eqnarray}\label{app_formula_cm_last1}
    \beta(g)=\sum\limits_{k=0}^{\infty}\beta_k g^k, \ \ \beta_{loop}(g)=\beta(g)-g. 
\end{eqnarray}
Next, we resum the loop part of the $\beta$ function taking into account the 
asymptotic behavior of its coefficients. The resummation parameter $b$, that is necessary for a better fitting of the $\beta$-function coefficients behaviour according to \eqref{lip_as}, varies within
$b_{range}=[0,25,0.5]$, $\lambda$ -- within $\lambda_{range}=[0,4.5,0.02]$. The parameter $\lambda$ allows us to also adjust the growth $B^{N}_{b}[\beta](g)$ from \eqref{pbl_proc} as $\sim g^{\lambda}$ that markedly  improves the convergence of numerical estimates.

We take the first point from $b_{range}\times \lambda_{range}$ and calculate resummed 
$\beta_{loop}(g)$ by means of the following formulas:
\begin{gather}\label{app_formula_cm_last2}
     \beta_{loop}(g)\approx\beta_{b,\lambda,loop}^{N}(g)=\int\limits_{0}^{\infty}dt \ t^b e^{-t}\left(\frac{gt}{w(gt)}\right)^{\lambda}\sum\limits_{k=0}^{N}W_{k,b,\lambda}[\beta_{loop}](w(gt))^k, \\ \nonumber
    \left(\frac{g}{w(g)}\right)^{\lambda}\sum\limits_{k=0}^{N}W_{k,b,\lambda}[\beta_{loop}](w(g))^k=\sum\limits_{k=0}^{N}\frac{\beta_k}{\Gamma(k+b+1)}g^k+\bigo{g^{N+1}}, \quad   w(g)=\frac{\sqrt{1+ag}-1}{\sqrt{1+ag}+1}.
\end{gather}
According to the asymptotic analysis, $a$ defines the closest to the origin singularity as -$1/a$ of Borel-transformed  function $B^{N}_{b}[\beta](g)$ in \eqref{pbl_proc}.

Having obtained $\beta_{b,\lambda,loop}^{N}$ 
as function of $g$, we have to find a minimal value of the following functional on 
some a~priori known grid, for example, from the PBL analysis with the step $\Delta g$: 
\begin{equation}
    F[g]=\big|\beta_{b,\lambda,loop}^{N}(g)+g\big|, \  \ \ g^*:  \ F[g^*]=\min\limits_{g}\big\{F[g]\big\}.
\end{equation}

\begin{table}
\caption{The Wilson fixed point location obtained from six--loop RG   
expansion for $\beta$ function. For $n > 5$ the values of $g^*$ given 
by CB and PBL resummations practically coincide to each other. 
The estimates found earlier using Pad\'e--Borel (PB), Pad\'e--Borel--Leroy 
(PBL) and Borel--Leroy + conformal mapping (CB) resummation procedures 
are also presented for comparison.}
\label{tab1}
\begin{center}
\begin{tabular}{c|c|c|c|c|c|c|c}
\hline
$n$ & CB ($n \le5$), & PB \cite{PhysRevE.51.1894} & PBL \cite{PhysRevB.17.1365} &CB \cite{PhysRevLett.39.95} & CB \cite{GZJ98} & PBL \cite{S98} & CB \cite{BT2005}     \\

    & PBL ($ n > 5 $), & & & & & &     \\
    
    & this work   & & & & & &     \\ 
\hline
0   & 1.4185    & 1.402  & 1.421    &1.417  & 1.413(6) &        &        \\
\hline
1   & 1.4166    & 1.401  & 1.416   &1.414  & 1.411(4) & 1.419  &        \\
\hline
2   & 1.4061    & 1.394  & 1.406    &1.405  & 1.403(3) & 1.4075 &        \\
\hline
3   & 1.3914    & 1.383  & 1.392      &1.391  & 1.390(4) & 1.392  &        \\
\hline
4   & 1.3745    & 1.369  &           &      & 1.377(5) & 1.3745 &        \\
\hline
5   & 1.3566    & 1.353  &           &     &          & 1.3565 & 1.3569 \\
\hline
6   & 1.3388    & 1.336  &        &      &          & 1.3385 & 1.3397 \\
\hline
8   & 1.3050    & 1.303  &        &         &          & 1.3045 &        \\
\hline
10  & 1.2754    & 1.274  &        &        &          & 1.2745 &        \\
\hline
12  & 1.2491    & 1.248  &        &        &          & 1.2487 &        \\
\hline
14  & 1.2270    & 1.226  &        &        &          & 1.2266 &        \\
\hline
16  & 1.2080    & 1.207  &        &        &          & 1.2077 &        \\
\hline
18  & 1.1918    & 1.191  &        &        &          & 1.1914 &        \\
\hline
20  & 1.1776    & 1.177  &        &        &          & 1.1773 &        \\
\hline
24  & 1.1545    & 1.154  &        &        &          & 1.1542 &        \\
\hline
28  & 1.1365    & 1.136  &        &        &          & 1.1361 &        \\
\hline
32  & 1.1222    & 1.122  &        &        &          & 1.1218 & 1.1219 \\
\hline
48  & 1.0858    &        &        &        &          &        &        \\
\hline
64  & 1.0659    &        &        &        &          &        & 1.0656 \\
\hline
\end{tabular}
\end{center}
\end{table}
If this minimal value is bigger than some marginal one -- $\varepsilon_{\beta}$, which 
we set at the beginning of the procedure \footnote{If the functional takes smaller values 
than $\varepsilon_{\beta}$ we assume that we found a zero of $\beta$ function.}, we 
discard currently considered point $(b,\lambda)$ and move to the next one, otherwise 
we begin to analyze the stability of the result obtained. 
In fact, the found candidate for the fixed point coordinate is a function of the 
resummation parameters -- $g^*(b,\lambda)$. As was already claimed, these parameters 
have no physical meaning, therefore it looks reasonable to adopt as optimal such their 
values in the vicinity of which the estimate would be least sensitive to their variations. 
To do this, we use proposed in \cite{KP17} measure of spreading (error bar) -- 
$\Delta_{s}(b,\lambda)$. Having ended the first cycle, we add all information related to 
this point -- $\{(b,\lambda),F[g^*],g^*,\Delta_{s}(b,\lambda)\}$ -- to the database 
and then repeat all the above steps for the whole grid of resummation parameters.

\begin{figure}
\begin{subfigure}{.5\textwidth}
\centering
\includegraphics[scale=0.5]{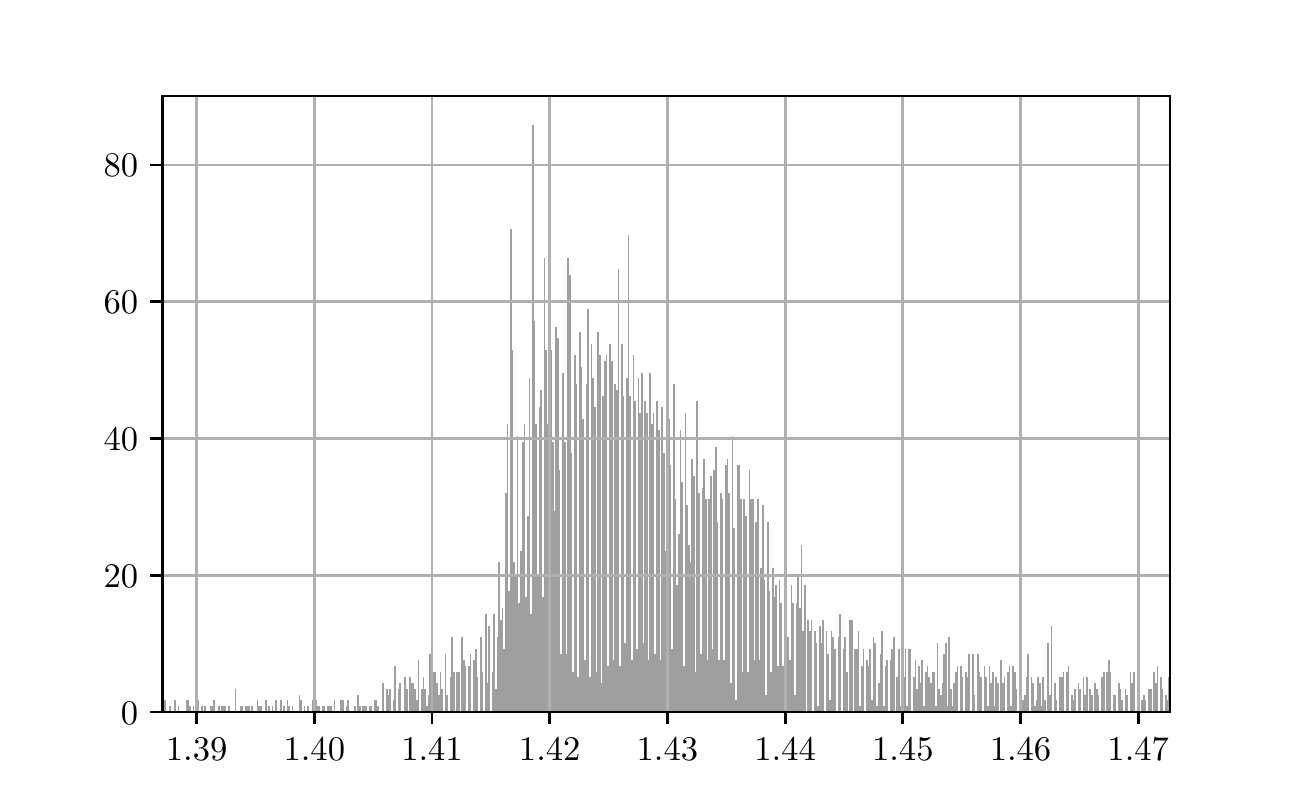}
\caption{For $n=0$: $g^*=1.4185$.}
\label{cm_fp_small_hist_0}
\end{subfigure}
\begin{subfigure}{0.5\textwidth}
\centering
\includegraphics[scale=0.5]{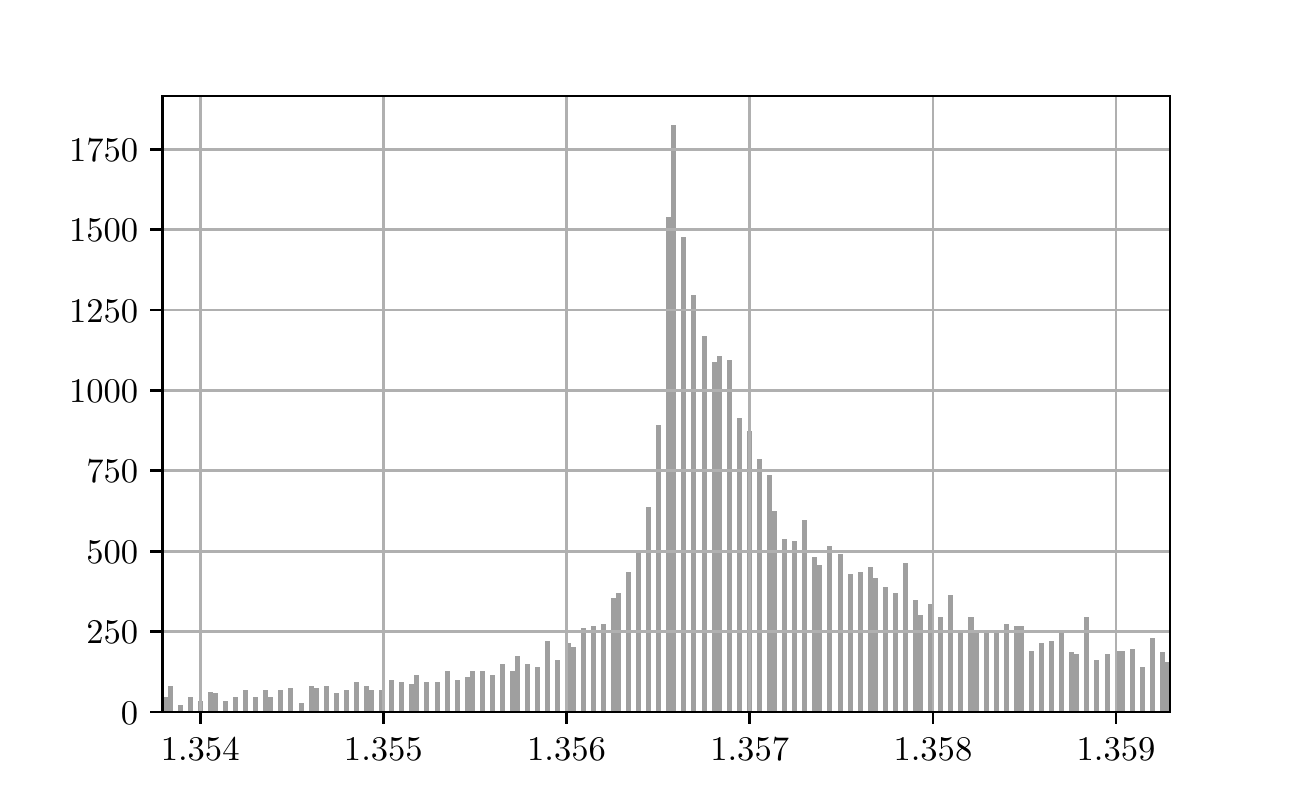}
\caption{For $n=5$: $g^*=1.3566$.}
\label{cm_fp_small_hist_5}
\end{subfigure}
\caption{Histograms of CB estimates for the fixed point coordinate for $n=0$ 
and $n=5$. The vertical axis represents the number of hits in a particular bin. 
The difference in the quality of pictures is caused by a different size of sample, 
each element of which has to meet certain requirements.}
\label{fp_cm_small_0_5_hist}
\end{figure}

The minimal value of the $\Delta_{s}(b,\lambda)$ is, in fact, an indicator that analyzed 
function -- $\beta_{b,\lambda,loop}^{N}(g^*)$ --  achieved a plateau. After analyzing 
the whole grid, we find point through collected data with the minimal value of 
$\Delta_{s}(b^*,\lambda^*)$. It is natural to take as a final estimate of the fixed point 
coordinate the corresponding value of $g^*(b^*,\lambda^*)$. In order to exclude the 
possibility of accidentally extreme deviations we perform the following steps. We 
construct the set of points, the spreading of which would not exceed the minimal one -- 
$\Delta_{s}(b^*,\lambda^*)$ -- by more than three times. Then we take the weighted 
average of the obtained sample and tripled standard deviation is accepted as an error bar. 
To illustrate the work of this algorithm, the histograms for $n = 0$ and $n = 5$ with 
distribution of $g^*(b,\lambda)$ constructed on the base of the whole parameter grid are 
presented in Figure \ref{fp_cm_small_0_5_hist}. 

The values of $g^*$ corresponding to picks of histograms are accepted as final 
estimates of the Wilson fixed point location given by CB resummation procedure. 
They are collected, along with PBL estimates, in Table \ref{tab1}.    

\section{Octic effective interaction at criticality}

To get an idea about numerical structure of the expansions for the octic
effective coupling let us consider the series obtained under physically
most interesting values $n = 1$ and $n = 3$ corresponding to simple
fluids, binary mixtures, uniaxial and Heisenberg ferromagnets etc. It is
instructive also to address the cases $n = 10$ (superfluid neutron liquid
\cite{SS78,S80n}), $n = 18$ (superfluid helium-3 \cite{MS73,S80h}) and $n
= 64$ that shed light on the behavior of the RG and pseudo-$\varepsilon$
expansions in the limit $n \gg 1$. These series read:

\begin{equation}
\begin{split}
R_8=&-9 g \left(1 - 2.4074074 g
+3.7894413 g^2 - 6.3233302 g^3\right), \ \ \ n = 1 \\
R_8=&-\frac{83}{11}g \left(1 - 2.1233892 g
+ 2.8877222 g^2 - 4.1661684 g^3 \right), \ \ \ n = 3 \\
R_8=&-5 g \left(1 - 1.6323731 g
+ 1.5565223 g^2 - 1.5739367 g^3 \right), \ \ \ n = 10 \\
R_8=&-\frac{49}{13} g \left(1 - 1.4015156 g
+ 1.0224677 g^2 - 0.7939063 g^3\right), \ \ \ n = 18 \\
R_8=&-2 g \left(1 - 1.0979081 g
+ 0.3632743 g^2 -0.1382065 g^3 \right), \ \ \ n = 64, \\
\end{split}
\end{equation}

\begin{equation}
\begin{split}
g_8^*=&-\frac{8\pi^3}{81}\tau^4 \left(1 - 0.7174211 \tau
- 0.2013970 \tau^2 - 0.706239\tau^3\right), \ \ \ n=1 \\
g_8^*=&-\frac{664\pi^3}{14641}\tau^4 \left(1 - 0.5904844\tau
- 0.2566839\tau^2 - 0.446149\tau^3\right), \ \ \ n=3 \\
g_8^*=&-\frac{5\pi^3}{729}\tau^4 \left(1 - 0.5349794\tau
- 0.2352017\tau^2 - 0.141469\tau^3\right), \ \ \ n=10 \\
g_8^*=&-\frac{49\pi^3}{28561}\tau^4 \left(1 - 0.5880157\tau
- 0.1936037 \tau^2 - 0.051025\tau^3\right), \ \ \ n=18 \\
g_8^*=&-\frac{\pi^3}{23328}\tau^4 \left(1 - 0.7762346\tau
- 0.0866804\tau^2 + 0.014012\tau^3\right), \ \ \ n=64, \\
\end{split}
\end{equation}

\begin{equation}
\begin{split}
R_8^*=&-9\tau \left(1 - 1.9849108 \tau + 1.7611357 \tau^2
- 1.966585 \tau^3\right), \ \ \ n=1 \\
R_8^*=&-\frac{83}{11}\tau \left(1 - 1.7401630\tau + 1.2710234\tau^2
- 1.194256\tau^3\right), \ \ \ n=3 \\
R_8^*=&-5\tau \left(1 - 1.3580247 \tau + 0.6598114 \tau^2
- 0.409513 \tau^3\right), \ \ \ n=10 \\
R_8^*=&-\frac{49}{13}\tau \left(1 - 1.1981406 \tau + 0.4426329 \tau^2
- 0.201501 \tau^3\right), \ \ \ n=18 \\
R_8^*=&-2\tau \left(1 - 1.0174897 \tau + 0.1748660 \tau^2
- 0.033612 \tau^3\right), \ \ \ n=64. \\
\end{split}
\end{equation}

As is seen, the structure of $\tau$-series for $R_8^*$ is more favorable
from the numerical point of view than that for the coupling constant.
Indeed, although the coefficients of the series (26) are bigger than their
counterparts for $g_8^*$, these series are alternating, i.e. have a
regular structure. Moreover, the universal ratio $R_8$ itself enters the
expressions for free energy and nonlinear susceptibilities determining
important physical quantities. That is why further we will work with the
series for this universal ratio.

\begin{figure}
\begin{subfigure}{.5\textwidth}
\centering
\includegraphics[scale=0.56]{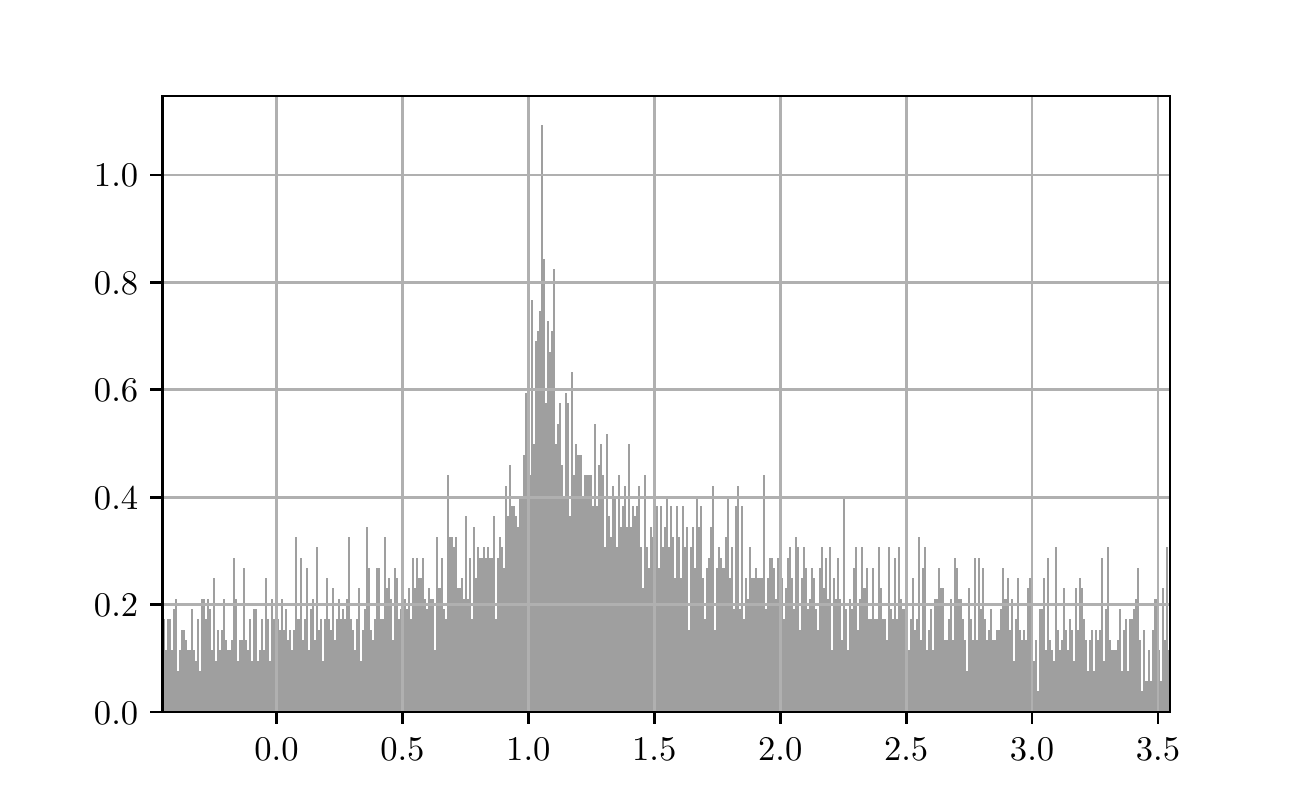}
\caption{$n=0$: $R_8=1.3698$.}
\label{cm_ce_r8_0_hist_grid}
\end{subfigure}
\begin{subfigure}{.5\textwidth}
\centering
\includegraphics[scale=0.56]{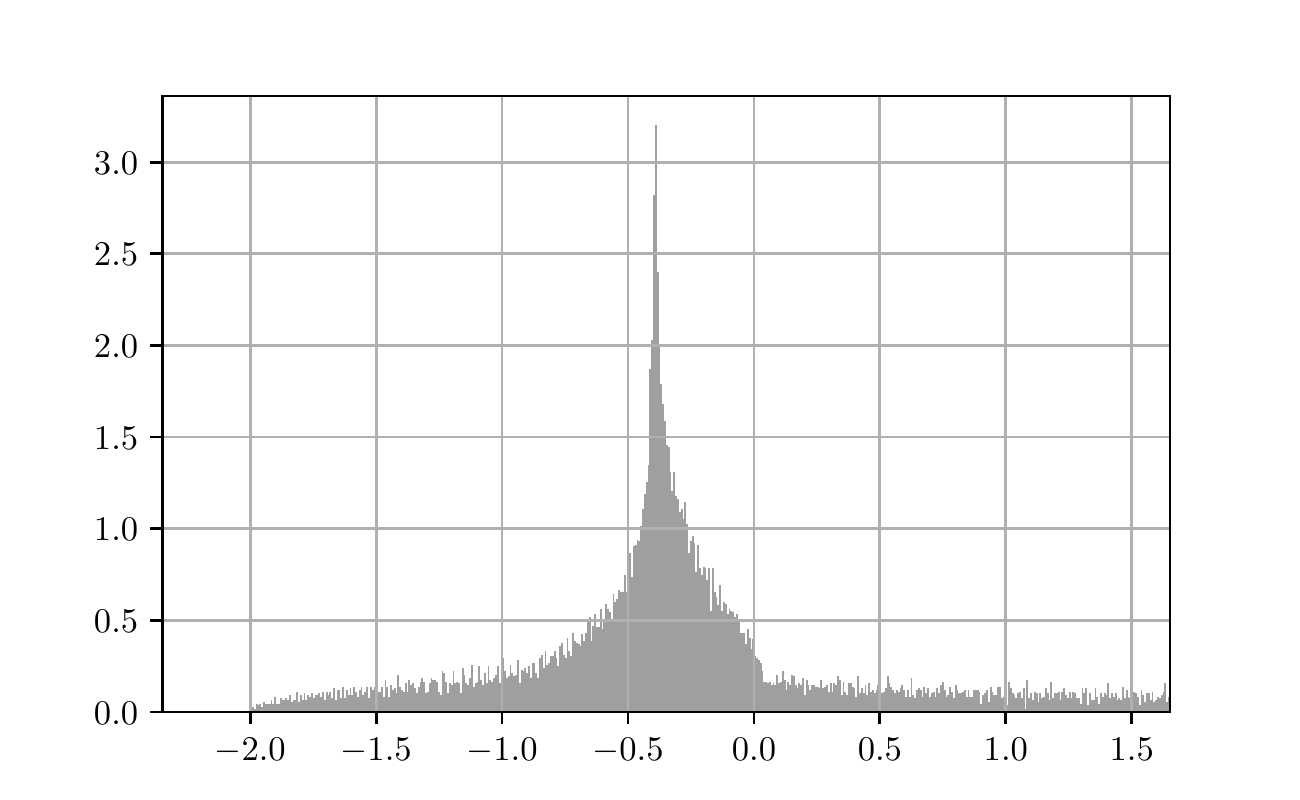}
\caption{$n=12$: $R_8=-0.3739$.}
\label{cm_ce_r8_12_hist_grid}
\end{subfigure}
\caption{Distributions of numerical estimates for universal ratio $R_8^*$ given by CB resummed 
3D RG expansions that are generated by the variation of resummation parameters. Here the results 
are presented for $n=\{0,12\}$. Others can be found in Supplementary materials to this work. 
The vertical axis represents the frequency of hits in a particular bin. The difference in the 
quality of pictures is caused by peculiar behavior of the series at a specific value of $n$ 
and also by different sizes of sample each element of which has to meet certain requirements.}
\label{r8_rg_cm_small_art_0_12}
\end{figure}

\begin{figure}
\begin{subfigure}{.5\textwidth}
\centering
\includegraphics[scale=0.52]{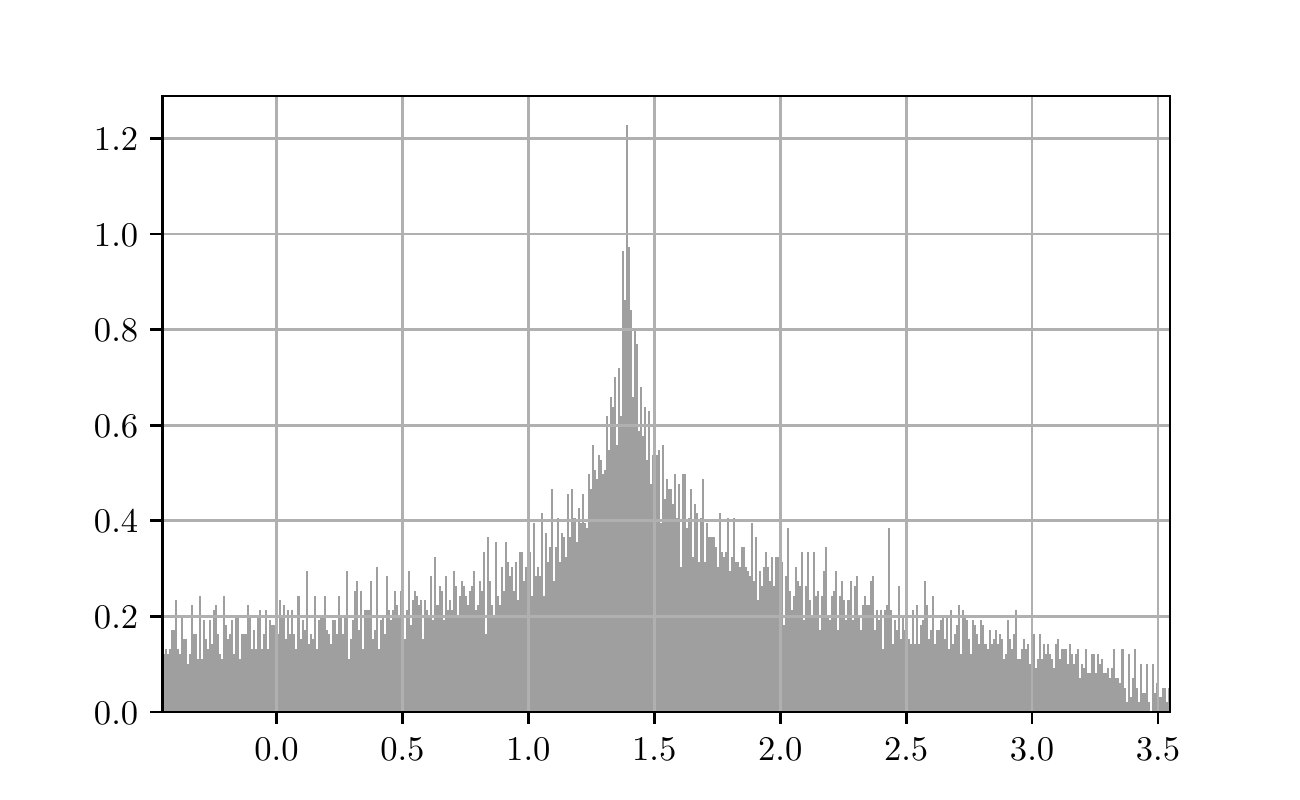}
\caption{For $n=0$: $R_8=1.6484$.}
\label{cm_ce_r8_0_hist_grid_pe}
\end{subfigure}
\begin{subfigure}{.5\textwidth}
\centering
\includegraphics[scale=0.52]{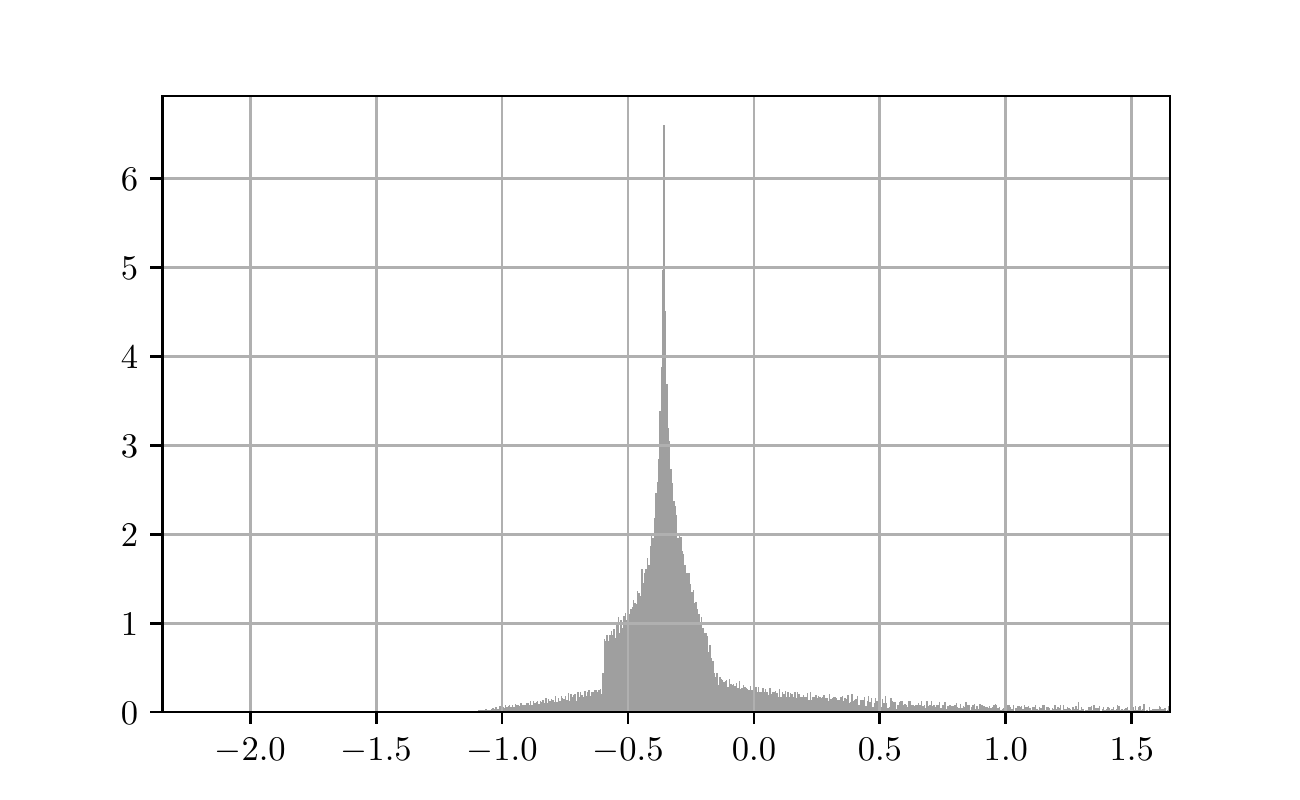}
\caption{For $n=12$: $R_8=-0.3430$.}
\label{cm_ce_r8_12_hist_grid_pe}
\end{subfigure}
\caption{Distributions of CB estimates resulting from pseudo-$\varepsilon$ expansion for universal 
ratio $R_8^*$ generated by the variation of resummation parameters. The results for $n = 0$ and $n = 12$
are presented. Others -- for different values of $n$ -- can be found in Supplementary materials. 
The vertical axis represents the frequency of hits in a particular bin. The difference in the quality 
of pictures is caused by peculiar behavior of $\tau$-series at a specific value of $n$ and also by 
different sizes of sample each element of which has to meet certain requirements.}
\label{r8_pe_cm_small_art_0_12}
\end{figure}

Since the series (24), (26) are divergent, to get proper numerical 
estimates as usual we have to apply resummation procedures. Let us first process the 
RG expansion and $\tau$-series for $R_8^*$ by means of PBL resummation 
technique. By analogy with the calculation of the fixed point the value 
of the parameter $b$ will be referred to as optimal, 
$b_{opt}$, if it provides the fastest convergence of the series, i. e. 
minimizes the differences between the estimates given by the most stable 
(diagonal or near-diagonal) Pad\'e approximants for the Borel-Leroy image. 
Looking for more advanced resummation technique we address the CB machinery. 
The procedure is similar to that used for evaluating the Wilson fixed point 
location with an important simplification -- there is no necessity in the 
coordinate space probing. The main working formula reads 
\begin{equation}\label{app_formula_cm_last}
    f(g)\approx f^{(N)}_{b,\lambda}(g) = \int\limits_{0}^{\infty}dt \ t^b e^{-t}\left(\frac{gt}{w(gt)}\right)^{\lambda}\sum\limits_{k=0}^{N}W_{k,b,\lambda}(w(gt))^k,
\end{equation}
where coefficients are defined in the same way as in (\ref{app_formula_cm_last2}). 
The choice of fitting parameters is determined by the region of greatest stability, 
which, in turn, is determined by the special error function suggested in \cite{KP17}. 
The probing range for $b$ and $\lambda$ is the same as in the case of the fixed point 
coordinate search. A couple of histograms with the distribution of the universal ratio 
values under the whole grid of the resummation parameters are shown in Figures \ref{r8_rg_cm_small_art_0_12} and \ref{r8_pe_cm_small_art_0_12} for 3D RG and pseudo-$\varepsilon$ expansions respectively.

\begin{figure}[!ht]
\centering
\includegraphics[scale=0.82]{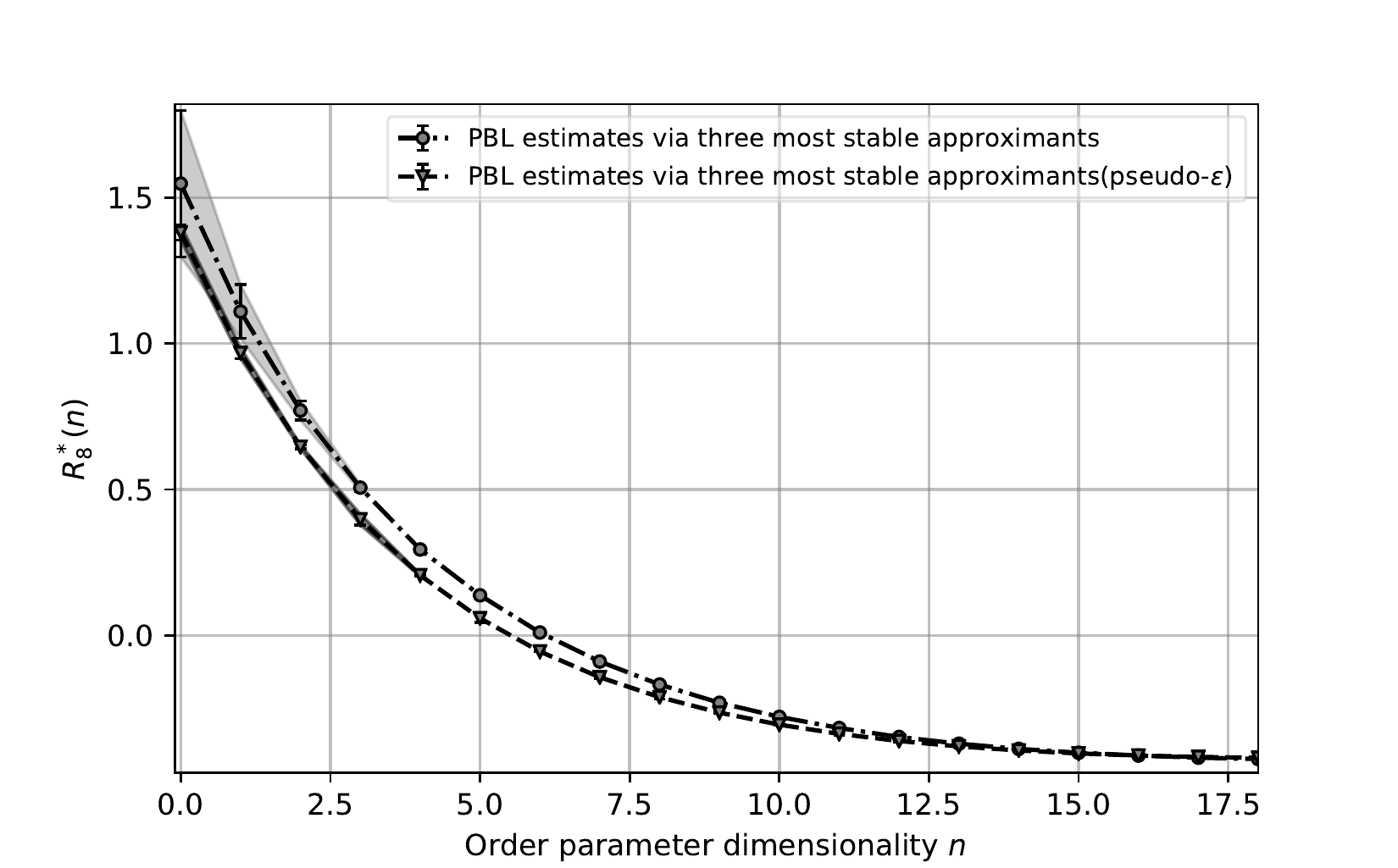}
\caption{PBL estimates resulting from pseudo-$\varepsilon$ and 3D RG expansions 
of universal ratio $R_8^*$ for different values of $n$.}
\label{pbl_0_64_r8}
\end{figure}
Four sets of $R_8^*$ estimates obtained from four-loop RG and pseudo-$\varepsilon$ 
expansions within PBL and CB resummation approaches are collected in Table \ref{tab2}, along with the low-order RG, $\varepsilon$ expansion and lattice estimates found earlier. 
PBL estimates are also depicted in Figure \ref{pbl_0_64_r8} to give a general view 
of $R_8^*$ as function of $n$. 

\begin{table}[!ht]
\caption{The values of $R_8^*$ for various $n$ found by resummation of 
the four-loop (4l) RG series and of corresponding pseudo-$\varepsilon$ expansion 
(P$\varepsilon$E). The estimates of $R_8^*$ resulting from the three-loop (3l) RG 
series in 3 dimensions \cite{GZ97,SOUK99,BT2005}, obtained within the $\varepsilon$ 
expansion ($\varepsilon$ exp) approach \cite{GZ97,PV2000,BT2005} and extracted from 
lattice calculations (LC) are also presented for comparison.}\label{tab2}
\begin{center}
\begin{tabular}{*{8}{c}}
\hline
$n$ & 4l RG & 4l P$\varepsilon$E & 4l RG & 4l P$\varepsilon$E & 3l RG & $\varepsilon$ exp & LC \\
    & PBL    & PBL  & CB & CB &  PBL \cite{SOUK99} & \cite{PV2000} & \\
\hline
0 & ~~1.548  & ~~1.382  & ~~1.370 & ~~1.648 & ~~  & ~1.1(2)~ & ~~  \\
1 & ~~1.109  & ~~0.968  & ~~0.967 & ~~1.159 & ~0.856~ & ~0.94(14)~ & ~0.871(14)\cite{BP11}~  \\
  &     &    &  &  & ~0.857(86)\cite{GZ97}~ & ~0.78(5)\cite{GZ97}~ & 0.79(4)\cite{CPRV2002} \\
2 & ~~0.770  & ~~0.646  & ~~0.656 & ~~0.786 & ~0.563~ & ~0.71(16) & 0.494(34)\cite{CHPRV2001} \\
3 & ~~0.506  & ~~0.397  & ~~0.412 & ~~0.503 & ~0.334~ & ~0.33(10) & 0.21(7)\cite{CHPRV2002} \\
4 & ~~0.299  & ~~0.206  & ~~0.222 & ~~0.286 & ~0.15~  & ~0.065(80) & 0.07(14)\cite{TPV03} \\
5 & ~~0.137  & ~~0.058  & ~~0.063 & ~~0.123 & $-$0.3(9)\cite{BT2005} & $-$0.1(2)\cite{BT2005} & \\
6 & ~~0.010  & $-$0.056 & $-$0.051 & $-$0.004 & $-$0.09 & $-$0.2(1)\cite{BT2005} & \\
7 & $-$0.090 & $-$0.144 & $-$0.140 & $-$0.101 &          &               & \\
8 & $-$0.169 & $-$0.212 & $-$0.213 & $-$0.177 & $-$0.25  & $-$0.405(31) & \\
10 & $-$0.280 & $-$0.306 & $-$0.313 & $-$0.281 &         &               & \\
16 & $-$0.412 & $-$0.412 & $-$0.428 & $-$0.400 & $-$0.44 & $-$0.528(14) & \\
24 & $-$0.424 & $-$0.413 & $-$0.431 & $-$0.408 &         &               & \\
32 & $-$0.394 & $-$0.380 & $-$0.396 & $-$0.377 & $-$0.42  & $-$0.425(7)  & \\
   &          &          &  &  & $-$0.45(7)\cite{BT2005} & $-$0.427(3)\cite{BT2005} & \\
48 & $-$0.324 & $-$0.311 & $-$0.324 & $-$0.310 & $-$0.35~ & $-$0.322(2)  &   \\
64 & $-$0.270 & $-$0.259 & $-$0.269 & $-$0.259 & $-$0.29(3)\cite{BT2005} & $-$0.269(3)\cite{BT2005} & \\ 
\hline
\end{tabular}
\end{center}
\end{table}
At the end of this section let us make a few remarks. It is worthy to note that errors 
of PBL estimates extracted from the corresponding pseudo-$\varepsilon$ expansions within the chosen resummation strategy have been severely underestimated, especially for 
small $n$. Generally speaking, the estimating of errors for quantities obtained from 
divergent series is a tricky problem solution of which is often based on some empirical 
judgments and experience. Therefore these error estimates can not be treated as fully reliable. 
As for the values of $R_8^*$ themselves, one can see that with growing $n$ the numerical 
estimates become more and more stable with respect to the resummation approach used. 
At the same time, for physically interesting cases $n=0,1,2,3$ the situation remains 
somewhat dramatic even in the four-loop approximation. Indeed, the values of $R_8^*$ 
given by 3D RG and pseudo-$\varepsilon$ expansions are scattered within big (15-25\%) intervals and differ considerably from the three-loop RG estimates. This reflects the 
unfavorable structure of the RG series and pseudo-$\varepsilon$ expansion for $R_8^*$ 
discussed in Introduction and signals that higher-order RG calculations of this quantity are 
certainly desirable. Such calculations, however, are expected to be rather complicated 
since finding of the next, five-loop contribution to the universal eight-order coupling 
will require evaluation of 2319 Feynman graphs.          

\section{Tenth-order coupling in the critical region}

To analyze the structure of RG expansions for $R_{10}$ and $\tau$-series 
for $R_{10}^*$, we proceed in the same way as in the case of the eight-order 
coupling. For selected values of $n$ the series of interest are:

\begin{equation}
\begin{split}
R_{10}=&54 g \left(1 - 4.4444444 g
+ 11.299686 g^2\right), \ \ \ n = 1 \\
R_{10}=&\frac{490}{11}g \left(1 - 3.8586866 g
+ 8.4785734 g^2 \right), \ \ \ n = 3 \\
R_{10}=&56 g \left(1 - 5.6888105 g
+ 17.951828 g^2 \right), \ \ \ n = 10 \\
R_{10}=&\frac{260}{13} g \left(1 - 2.3672876031 g
+ 2.994454005 g^2\right), \ \ \ n = 18 \\
R_{10}=&\frac{17}{2}g \left(1 - 1.7610546276 g
+ 1.3767008404908 g^2\right), \ \ \ n = 64, \\
\end{split}
\end{equation}

\begin{equation}
\begin{split}
g_{10}^*=&\frac{32\pi^4}{243}\tau^5 \left(1 - 2.3319616 \tau +
1.84782991 \tau^2\right), \ \ \ n=1 \\
g_{10}^*=&\frac{7840\pi^4}{161051}\tau^5 \left(1 - 1.9425556\tau
+1.1285875 \tau^2\right), \ \ \ n=3 \\
g_{10}^*=&\frac{28\pi^4}{6561}\tau^5 \left(1 - 1.4726631 \tau
+ 0.55331470\tau^2 \right), \ \ \ n=10 \\
g_{10}^*=&\frac{20\pi^4}{28561}\tau^5 \left(1 - 1.3504127\tau
+0.47054348\tau^2\right), \ \ \ n=18 \\
g_{10}^*=&\frac{17\pi^4}{3359232}\tau^5 \left(1 - 1.3589627\tau
+0.53252324\tau^2\right), \ \ \ n=64, \\
\end{split}
\end{equation}

\begin{equation}
\begin{split}
R_{10}^*=&54\tau \left(1 - 4.0219479 \tau + 7.5500981 \tau^2
\right), \ \ \ n=1 \\
R_{10}^*=&\frac{490}{11}\tau \left(1 - 3.4754604 \tau
+ 5.5318517 \tau^2\right), \ \ \ n=3 \\
R_{10}^*=&28\tau \left(1 - 2.5700568 \tau + 2.92620786 \tau^2
\right), \ \ \ n=10 \\
R_{10}^*=&20\tau \left(1 - 2.1639126 \tau +  2.0217915 \tau^2
\right), \ \ \ n=18 \\
R_{10}^*=&\frac{17}{2}\tau \left(1 - 1.6806362 \tau
+ 1.0816342 \tau^2\right), \ \ \ n=64. \\
\end{split}
\end{equation}

These series possess big coefficients even for $n \gg 1$, to say nothing about those for physical values $n = 1, 2, 3$. It is not surprising, therefore, that application of all the resummation techniques having used above leads to quite chaotic estimates for 
$R_{10}^*$ covering huge (several dozens) intervals. Of course, the series found 
are too short to provide accurate estimates but their small length is not the only reason of such a failure. Rather strong divergence of 3D RG expansion for universal tenth-order 
coupling seems to play more important role. Considerable scattering of numerical estimates 
for $R_{10}^*$ obtained in the five-loop RG approximation for 3D Ising model ($n = 1$) 
\cite{SNK17} confirms this conclusion. This implies that calculation of higher-order terms 
in corresponding RG expansion may turn out to be insufficient for getting proper numerical 
results. Moreover, such a calculation would be really complicated since it includes 
evaluation of thousands of graphs\footnote{For example, in order to obtain four-loop contribution 
to $R_{10}$ one should calculate 684 Feynman diagrams.}. On the other hand, working 
with lengthy enough RG series may help to overcome the problem of its dramatic divergence and 
result in fair numerical estimates.         

\section{Conclusion}

To summarize, for 3D $O(n)$-symmetric $\lambda\phi^4$ field theory we 
have calculated the four-loop RG contribution to the universal ratio $R_8$ 
and three-loop RG series for $R_{10}$. This has required to evaluate in total 
339 Feynman diagrams. We have also found corresponding four-loop and three-loop 
pseudo-$\epsilon$ expansions. The series for $R_8$ have been resummed by means 
Pade--Borel--Leroy and conform--Borel resummation techniques and calculated 
additive has been found to appreciably shift the numerical estimates for this 
coupling under the physical values of $n$. On the other hand, for these $n$ the 
numerical estimates for $R_8$ have turned out to notably depend on the resummation 
procedure signaling that higher-order RG calculations of this quantity remain 
desirable. Three-loop RG and pseudo-$\epsilon$ expansions for $R_{10}$ have been 
found to strongly diverge both for physical $n$ and $n \gg 1$: they possess big 
coefficients $C_k$ that grow or, at least, do not regularly diminish with growing $k$. 
This prevents extracting numerical values of this universal ratio from the series 
found. Perhaps, the calculation of rather lengthy RG expansions for $R_{10}$ would 
soften this problem and pave the way to getting proper numerical estimates.       

\section*{Acknowledgment}
AK gratefully acknowledges the support of the Foundation for Advancement of 
Theoretical Physics "BASIS" under grant 18-1-2-43-1.

\appendix
\section{Supplementary materials }
\label{app:suppl}
In Supplementary materials we present one-particle irreducible Feynman graphs forming RG expansions for $g_8$ and $g_{10}$ couplings(\textit{rg\_expansions\_couplings\_g8\_g10.pdf}). It includes integrals, symmetry and tensor factors. The Nickel's notations describing graphs topology are used there.
\newpage 

\bibliographystyle{elsarticle-num}
\bibliography{houc}

\end{document}

\title{Universal effective couplings of the three-dimensional
$n$-vector model and field theory}

\author[label1]{\corref{cor2}A.\,Kudlis}
\ead{andrewkudlis@gmail.com}\cortext[cor2]{Corresponding author}

\author[label1]{A.\,I.\,Sokolov}

\address[label1]{St. Petersburg State University, 7/9 Universitetskaya Emb., St. Petersburg, 199034 Russia}

\date{\today}

\begin{abstract}
This document contains Tables with coefficients of RG expansions for $g_8$ and $g_{10}$  in four- and three-loop approximation respectively.
\end{abstract}

\begin{keyword}
renormalization group, multi-loop calculations, effective coupling
constants, universal ratios.

\MSC{82B28}
\end{keyword}

\maketitle

\section{Tables}
\begin{center}
\renewcommand*{\arraystretch}{1.25}
\begin{longtable}{|l|l|l|l|l|}
\caption[Feynman graphs for effective coupling $g_8$]
{Feynman graphs for effective coupling $g_8$} 
\label{tab-a1} \\
\hline \multicolumn{1}{|c}{\textbf{No.}} &
\multicolumn{1}{|c}{\textbf{Graph description,}} &
\multicolumn{1}{|c}{\textbf{Integral value $\times$ $(8\pi)^L$,}} &
\multicolumn{1}{|c}{\textbf{Symmetry}} &
\multicolumn{1}{|c|}{\textbf{Tensor factor $\times$ $3^{L+3}$,}} \\
\multicolumn{1}{|c}{\textbf{}} &
\multicolumn{1}{|c}{\textbf{Nickel's
notations}} &
\multicolumn{1}{|c}{\textbf{$L$ -- number of loops}}&
\multicolumn{1}{|c}{\textbf{factor}} &
\multicolumn{1}{|c|}{\textbf{$L$ --
number of loops}} \\
\hline
\endfirsthead

\multicolumn{5}{c}%
{\tablename\ \thetable{} Feynman graphs for effective coupling $g_8$} \\
\hline \multicolumn{1}{|c}{\textbf{No.}} &
\multicolumn{1}{|c}{\textbf{Graph description,}} &
\multicolumn{1}{|c}{\textbf{Integral value $\times$ $(8\pi)^L$,}} &
\multicolumn{1}{|c}{\textbf{Symmetry}} &
\multicolumn{1}{|c|}{\textbf{Tensor factor $\times$ $3^{L+3}$,}} \\
\multicolumn{1}{|c}{\textbf{}} &
\multicolumn{1}{|c}{\textbf{Nickel's
notations}} &
\multicolumn{1}{|c}{\textbf{$L$ -- number of loops}} &
\multicolumn{1}{|c}{\textbf{factor}} &
\multicolumn{1}{|c|}{\textbf{$L$ --
number of loops}} \\
\hline
\endhead

\hline \multicolumn{5}{|r|}{{Continued on next page}} \\ \hline
\endfoot
\hline \hline
\endlastfoot
1-1 & $ee12|ee3|ee3|ee|$ & 1/8            & 315 & $80 + n$\\
    &                    &                &     &         \\
2-1 & $ee11|23|ee4|ee4|ee|$ & 1/8         &  630 & $212 + 30 n + n^2$\\
2-2 & $ee12|ee2|34|ee4|ee|$ & 1/16        &  315 & $224 + 18 n + n^2$\\
2-3 & $ee12|ee3|ee4|e44|e|$ & 19/162      & 1260 & $214 + 29 n$\\
2-4 & $ee12|ee3|e34|e4|ee|$ & 5/108       & 2520 & $230 + 13 n$\\
2-5 & $ee12|e34|e34|ee|ee|$ & 1/27        &  420 & $234 + 9 n$\\
    &                    &          &     &         \\
3-1 & $ee11|22|34|ee5|ee5|ee|$ & 1/8                           &  315 & $528 + 168 n + 32 n^2 + n^3$\\
3-2 & $ee11|23|ee3|45|ee5|ee|$ & 1/16                          &  630 & $584 + 128 n + 16 n^2 + n^3$\\
3-3 & $ee11|23|ee4|ee4|55|ee|$ & 1/8                          & 157.5 & $560 + 152 n + 16 n^2 + n^3$\\
3-4 & $ee11|23|ee4|ee5|e55|e|$ & 19/162                        &  630 & $568 + 150 n + 11 n^2$\\
3-5 & $ee11|23|ee4|e45|e5|ee|$ & 5/108                         & 2520 & $608 + 114 n + 7 n^2$\\
3-6 & $ee11|23|ee4|e55|ee5|e|$ & 19/162                        & 1260 & $568 + 150 n + 11 n^2$\\
3-7 & $ee11|23|ee4|455|ee|ee|$ & 1/8                           &  315 & $560 + 152 n + 16 n^2 + n^3$\\
3-8 & $ee11|23|e34|e5|ee5|ee|$ & 5/108                         & 1260 & $600 + 118 n + 11 n^2$\\
3-9 & $ee11|23|e45|e45|ee|ee|$ & 1/27                          &  630 & $616 + 106 n + 7 n^2$\\
3-10 & $ee12|ee2|33|45|ee5|ee|$ & 1/16                        & 157.5 & $576 + 132 n + 20 n^2 + n^3$\\
3-11 & $ee12|ee2|34|ee5|e55|e|$ & 1/18                         & 1260 & $604 + 120 n + 5 n^2$\\
3-12 & $ee12|ee2|34|e45|e5|ee|$ & 1/36                         & 1260 & $636 + 88 n + 5 n^2$\\
3-13 & $ee12|ee3|ee3|45|e55|e|$ & 1/12                         &  630 & $584 + 142 n + 3 n^2$\\
3-14 & $ee12|ee3|ee4|ee5|555||$ & 415/4608 -- 5/16 log[4/3]    &  210 & $480 + 246 n + 3 n^2$\\
3-15 & $ee12|ee3|ee4|e45|55|e|$ & -- 71/864 + 89 log[4/3]/162  & 2520 & $588 + 140 n + n^2$\\
3-16 & $ee12|ee3|ee4|e55|e55||$ & 0.110545898524682            &  630 & $536 + 166 n + 27 n^2$\\
3-17 & $ee12|ee3|ee4|445|5|ee|$ & 31/384                       &  630 & $584 + 142 n + 3 n^2$\\
3-18 & $ee12|ee3|e34|e5|e55|e|$ & 17/108 -- 11 log[4/3]/27     & 2520 & $620 + 108 n + n^2$\\
3-19 & $ee12|ee3|e34|45|e5|ee|$ & 19/864                       & 5040 & $652 + 76 n + n^2$\\
3-20 & $ee12|ee3|e34|55|ee5|e|$ & 17/108 -- 11 log[4/3]/27     & 2520 & $616 + 110 n + 3 n^2$\\
3-21 & $ee12|ee3|e44|e45|5|ee|$ & 4 log[4/3]/27                & 2520 & $604 + 116 n + 9 n^2$\\
3-22 & $ee12|ee3|e44|e55|e5|e|$ & 0.110545898524682            & 1260 & $572 + 148 n + 9 n^2$\\
3-23 & $ee12|ee3|e45|e45|e5|e|$ & 0.019439866320025            & 2520 & $658 + 71 n$\\
3-24 & $ee12|ee3|e45|445|e|ee|$ & -- 47/288 + 19 log[4/3]/27   & 2520 & $620 + 108 n + n^2$\\
3-25 & $ee12|ee3|334|5|ee5|ee|$ & 13/256                       &  315 & $608 + 118 n + 3 n^2$\\
3-26 & $ee12|ee3|345|45|ee|ee|$ & 3/128                        &  630 & $648 + 78 n + 3 n^2$\\
3-27 & $ee12|e23|e4|ee5|e55|e|$ & 4 log[4/3]/27                & 5040 & $616 + 110 n + 3 n^2$\\
3-28 & $ee12|e23|e4|e45|e5|ee|$ & 1/54                         & 2520 & $660 + 68 n + n^2$\\
3-29 & $ee12|e23|45|ee4|e5|ee|$ & 1/54                         & 2520 & $656 + 70 n + 3 n^2$\\
3-30 & $ee12|e23|45|e45|ee|ee|$ & 7/432                        & 2520 & $668 + 60 n + n^2$\\
3-31 & $ee12|e33|e44|e5|e5|ee|$ & 0.110545898524682            &  630 & $572 + 148 n + 9 n^2$\\
3-32 & $ee12|e33|e45|45|ee|ee|$ & 11/54 -- 16 log[4/3]/27      & 1260 & $624 + 102 n + 3n^2$\\
3-33 & $ee12|e34|e34|e5|e5|ee|$ & 0.012797775151119            & 1260 & $682 + 47 n$\\
3-34 & $ee12|e34|e35|ee|e55|e|$ & 11/54 -- 16 log[4/3]/27      & 1260 & $628 + 100 n + n^2$\\
3-35 & $ee12|e34|e35|e4|e5|ee|$ & 0.014252593468145            & 5040 & $674 + 55 n$\\
3-36 & $ee12|345|345|ee|ee|ee|$ & 1/64                         & 52.5 & $672 + 54 n + 3 n^2$\\
     &                    &          &     &         \\
4-1  & $ee11|22|33|45|ee6|ee6|ee|$ & 1/8                      & 157.5 & $1264 + 656n + 232n^2 + 34n^3 + n^4$\\
4-2  & $ee11|22|34|ee4|56|ee6|ee|$ & 1/16                      &  315 & $1440 + 576n + 152n^2 + 18n^3 + n^4$\\
4-3  & $ee11|22|34|ee5|ee5|66|ee|$ & 1/8                      & 157.5 & $1392 + 624n + 152n^2 + 18n^3 + n^4$\\
4-4  & $ee11|22|34|ee5|ee6|e66|e|$ & 19/162                    &  315 & $1416 + 624n + 136n^2 + 11n^3$\\
4-5  & $ee11|22|34|ee5|e56|e6|ee|$ & 5/108                     & 1260 & $1512 + 552n + 116n^2 + 7n^3$\\
4-6  & $ee11|22|34|ee5|e66|ee6|e|$ & 19/162                    &  630 & $1416 + 624n + 136n^2 + 11n^3$\\
4-7  & $ee11|22|34|ee5|566|ee|ee|$ & 1/8                       &  315 & $1392 + 624n + 152n^2 + 18n^3 + n^4$\\
4-8  & $ee11|22|34|e45|e6|ee6|ee|$ & 5/108                     &  630 & $1480 + 560n + 136n^2 + 11n^3$\\
4-9  & $ee11|22|34|e56|e56|ee|ee|$ & 1/27                      &  315 & $1528 + 536n + 116n^2 + 7n^3$\\
4-10 & $ee11|23|ee3|44|56|ee6|ee|$ & 1/16                      &  315 & $1488 + 552n + 128n^2 + 18n^3 + n^4$\\
4-11 & $ee11|23|ee3|45|ee5|66|ee|$ & 1/16                      &  315 & $1520 + 544n + 108n^2 + 14n^3 + n^4$\\
4-12 & $ee11|23|ee3|45|ee6|e66|e|$ & 1/18                      & 1260 & $1576 + 532n + 74n^2 + 5n^3$\\
4-13 & $ee11|23|ee3|45|e56|e6|ee|$ & 1/36                      & 1260 & $1656 + 460n + 66n^2 + 5n^3$\\
4-14 & $ee11|23|ee4|ee4|56|e66|e|$ & 1/12                      &  315 & $1544 + 576n + 64n^2 + 3n^3$\\
4-15 & $ee11|23|ee4|ee5|ee6|666||$ & 415/4608 -- 5 log[4/3]/16 &  210 & $1272 + 816n + 96n^2 + 3n^3$\\
4-16 & $ee11|23|ee4|ee5|e56|66|e|$ & -- 71/864 + 89 log[4/3]/162 & 1260 & $1560 + 572n + 54n^2 + n^3$\\
4-17 & $ee11|23|ee4|ee5|e66|e66||$ & 0.110545898524682         &  315 & $1424 + 624n + 130n^2 + 9n^3$\\
4-18 & $ee11|23|ee4|ee5|556|6|ee|$ & 31/384                    &  315 & $1544 + 576n + 64n^2 + 3n^3$\\
4-19 & $ee11|23|ee4|e45|e5|66|ee|$ & 5/108                     & 1260 & $1584 + 520n + 78n^2 + 5n^3$\\
4-20 & $ee11|23|ee4|e45|e6|e66|e|$ & 17/108 -- 11 log[4/3]/27  & 2520 & $1640 + 500n + 46n^2 + n^3$\\
4-21 & $ee11|23|ee4|e45|56|e6|ee|$ & 19/864                    & 2520 & $1720 + 428n + 38n^2 + n^3$\\
4-22 & $ee11|23|ee4|e45|66|ee6|e|$ & 17/108 -- 11 log[4/3]/27  & 1260 & $1624 + 504n + 56n^2 + 3n^3$\\
4-23 & $ee11|23|ee4|e55|ee6|66|e|$ & 0.110545898524682         &  630 & $1424 + 624n + 130n^2 + 9n^3$\\
4-24 & $ee11|23|ee4|e55|e56|6|ee|$ & 4 log[4/3]/27             & 1260 & $1600 + 516n + 68n^2 + 3n^3$\\
4-25 & $ee11|23|ee4|e55|e66|e6|e|$ & 0.110545898524682         & 1260 & $1520 + 588n + 76n^2 + 3n^3$\\
4-26 & $ee11|23|ee4|e55|566|e|ee|$ & 19/162                    &  315 & $1504 + 592n + 86n^2 + 5n^3$\\
4-27 & $ee11|23|ee4|e56|ee5|66|e|$ & -- 71/864 + 89 log[4/3]/162 & 1260 & $1560 + 572n + 54n^2 + n^3$\\
4-28 & $ee11|23|ee4|e56|e55|6|ee|$ & 4 log[4/3]/27             & 1260 & $1600 + 516n + 68n^2 + 3n^3$\\
4-29 & $ee11|23|ee4|e56|e56|e6|e|$ & 0.019439866320025         & 2520 & $1744 + 414n + 29n^2$\\
4-30 & $ee11|23|ee4|e56|556|e|ee|$ & -- 47/288 + 19 log[4/3]/27 & 1260 & $1640 + 500n + 46n^2 + n^3$\\
4-31 & $ee11|23|ee4|445|6|ee6|ee|$ & 13/256                    &  630 & $1592 + 528n + 64n^2 + 3n^3$\\
4-32 & $ee11|23|ee4|455|e6|e6|ee|$ & 17/108 -- 11 log[4/3]/27  & 1260 & $1624 + 504n + 56n^2 + 3n^3$\\
4-33 & $ee11|23|ee4|455|66|ee|ee|$ & 1/8                       & 157.5 & $1472 + 592n + 108n^2 + 14n^3 + n^4$\\
4-34 & $ee11|23|ee4|456|ee|e66|e|$ & 1/12                      &  630 & $1544 + 576n + 64n^2 + 3n^3$\\
4-35 & $ee11|23|ee4|456|e5|e6|ee|$ & 19/864                    & 2520 & $1720 + 428n + 38n^2 + n^3$\\
4-36 & $ee11|23|ee4|456|56|ee|ee|$ & 3/128                     &  630 & $1704 + 432n + 48n^2 + 3n^3$\\
4-37 & $ee11|23|ee4|555|ee6|6|ee|$ & 415/4608 -- 5 log[4/3]/16 &  210 & $1272 + 816n + 96n^2 + 3n^3$\\
4-38 & $ee11|23|ee4|556|ee5|6|ee|$ & 31/384                    &  630 & $1544 + 576n + 64n^2 + 3n^3$\\
4-39 & $ee11|23|ee4|556|ee6|e6|e|$ & -- 71/864 + 89 log[4/3]/162 & 1260 & $1560 + 572n + 54n^2 + n^3$\\
4-40 & $ee11|23|ee4|556|e56|e|ee|$ & -- 47/288 + 19 log[4/3]/27 & 1260 & $1640 + 500n + 46n^2 + n^3$\\
4-41 & $ee11|23|ee4|556|e66|ee|e|$ & 19/162                    &  630 & $1504 + 592n + 86n^2 + 5n^3$\\
4-42 & $ee11|23|e34|e4|56|ee6|ee|$ & 1/36                      &  630 & $1648 + 468n + 68n^2 + 3n^3$\\
4-43 & $ee11|23|e34|e5|ee6|e66|e|$ & 4 log[4/3]/27             & 2520 & $1608 + 512n + 64n^2 + 3n^3$\\
4-44 & $ee11|23|e34|e5|e56|e6|ee|$ & 1/54                      & 2520 & $1720 + 420n + 46n^2 + n^3$\\
4-45 & $ee11|23|e34|45|e6|ee6|ee|$ & 19/864                    & 2520 & $1688 + 444n + 54n^2 + n^3$\\
4-46 & $ee11|23|e34|55|ee6|e6|ee|$ & 4 log[4/3]/27             & 1260 & $1560 + 524n + 94n^2 + 9n^3$\\
4-47 & $ee11|23|e34|56|ee5|e6|ee|$ & 1/54                      & 2520 & $1704 + 424n + 56n^2 + 3n^3$\\
4-48 & $ee11|23|e34|56|e56|ee|ee|$ & 7/432                     & 1260 & $1736 + 404n + 46n^2 + n^3$\\
4-49 & $ee11|23|e44|e45|6|ee6|ee|$ & 17/108 -- 11 log[4/3]/27  & 1260 & $1616 + 512n + 58n^2 + n^3$\\
4-50 & $ee11|23|e44|e55|e6|e6|ee|$ & 0.110545898524682         &  630 & $1520 + 588n + 76n^2 + 3n^3$\\
4-51 & $ee11|23|e44|e56|e5|e6|ee|$ & 4 log[4/3]/27             & 2520 & $1632 + 504n + 50n^2 + n^3$\\
4-52 & $ee11|23|e44|e56|56|ee|ee|$ & 11/54 -- 16 log[4/3]/27   &  630 & $1648 + 488n + 50n^2 + n^3$\\
4-53 & $ee11|23|e44|456|e|ee6|ee|$ & 1/18                      &  630 & $1584 + 532n + 68n^2 + 3n^3$\\
4-54 & $ee11|23|e45|e45|ee|66|ee|$ & 1/27                      &  315 & $1616 + 496n + 70n^2 + 5n^3$\\
4-55 & $ee11|23|e45|e45|e6|e6|ee|$ & 0.012797775151119         & 1260 & $1792 + 366n + 29n^2$\\
4-56 & $ee11|23|e45|e46|ee|e66|e|$ & 11/54 -- 16 log[4/3]/27   & 1260 & $1656 + 484n + 46n^2 + n^3$\\
4-57 & $ee11|23|e45|e46|e5|e6|ee|$ & 0.014252593468146         & 5040 & $1776 + 382n + 29n^2$\\
4-58 & $ee11|23|e45|e46|56|ee|ee|$ & 1/54                      & 1260 & $1728 + 416n + 42n^2 + n^3$\\
4-59 & $ee11|23|e45|446|e|ee6|ee|$ & -- 47/288 + 19 log[4/3]/27 & 1260 & $1624 + 508n + 54n^2 + n^3$\\
4-60 & $ee11|23|e45|456|ee|e6|ee|$ & 7/432                     & 2520 & $1752 + 396n + 38n^2 + n^3$\\
4-61 & $ee11|23|e45|466|ee|ee6|e|$ & 11/54 -- 16 log[4/3]/27   & 1260 & $1640 + 488n + 56n^2 + 3n^3$\\
4-62 & $ee11|23|e45|466|e5|ee|ee|$ & 5/108                     &  630 & $1600 + 512n + 70n^2 + 5n^3$\\
4-63 & $ee11|23|334|5|ee6|ee6|ee|$ & 31/384                    &  315 & $1480 + 608n + 96n^2 + 3n^3$\\
4-64 & $ee11|23|344|56|ee|ee6|ee|$ & 1/16                     & 157.5 & $1504 + 544n + 120n^2 + 18n^3 + n^4$\\
4-65 & $ee11|23|345|46|ee|ee6|ee|$ & 3/128                     &  630 & $1672 + 448n + 64n^2 + 3n^3$\\
4-66 & $ee11|23|456|456|ee|ee|ee|$ & 1/64                      &  105 & $1752 + 384n + 48n^2 + 3n^3$\\
4-67 & $ee12|ee2|33|44|56|ee6|ee|$ & 1/16                     & 78.75 & $1408 + 584n + 172n^2 + 22n^3 + n^4$\\
4-68 & $ee12|ee2|33|45|ee6|e66|e|$ & 1/18                      &  630 & $1560 + 540n + 82n^2 + 5n^3$\\
4-69 & $ee12|ee2|33|45|e56|e6|ee|$ & 1/36                      &  630 & $1624 + 476n + 82n^2 + 5n^3$\\
4-70 & $ee12|ee2|34|ee4|56|e66|e|$ & 1/24                      &  630 & $1616 + 516n + 52n^2 + 3n^3$\\
4-71 & $ee12|ee2|34|ee5|ee6|666||$ & 125/3456 -- log[4/3]/8    &  105 & $1344 + 780n + 60n^2 + 3n^3$\\
4-72 & $ee12|ee2|34|ee5|e56|66|e|$ & -- 1/48 + 7 log[4/3]/36   & 1260 & $1656 + 500n + 30n^2 + n^3$\\
4-73 & $ee12|ee2|34|ee5|e66|e66||$ & 0.050211596073066         &  630 & $1520 + 564n + 100n^2 + 3n^3$\\
4-74 & $ee12|ee2|34|ee5|556|6|ee|$ & 5/128                     &  630 & $1616 + 516n + 52n^2 + 3n^3$\\
4-75 & $ee12|ee2|34|ee5|566|e6|e|$ & -- 1/48 + 7 log[4/3]/36   & 1260 & $1656 + 500n + 30n^2 + n^3$\\
4-76 & $ee12|ee2|34|ee5|666|ee6||$ & 125/3456 -- log[4/3]/8    &  210 & $1344 + 780n + 60n^2 + 3n^3$\\
4-77 & $ee12|ee2|34|e45|e6|e66|e|$ & 1/18 -- log[4/3]/9        & 1260 & $1720 + 436n + 30n^2 + n^3$\\
4-78 & $ee12|ee2|34|e45|56|e6|ee|$ & 1/72                      & 2520 & $1784 + 372n + 30n^2 + n^3$\\
4-79 & $ee12|ee2|34|e45|66|ee6|e|$ & 1/18 -- log[4/3]/9        & 1260 & $1680 + 452n + 52n^2 + 3n^3$\\
4-80 & $ee12|ee2|34|e55|e56|6|ee|$ & 1/18 -- log[4/3]/9        & 1260 & $1688 + 452n + 46n^2 + n^3$\\
4-81 & $ee12|ee2|34|e55|e66|e6|e|$ & 0.050211596073066         &  630 & $1624 + 516n + 46n^2 + n^3$\\
4-82 & $ee12|ee2|34|e56|e56|e6|e|$ & 0.010868788166375         & 1260 & $1844 + 328n + 15n^2$\\
4-83 & $ee12|ee2|34|e56|556|e|ee|$ & -- 1/24 + 2 log[4/3]/9    & 1260 & $1720 + 436n + 30n^2 + n^3$\\
4-84 & $ee12|ee2|34|445|6|ee6|ee|$ & 5/128                     &  315 & $1600 + 524n + 60n^2 + 3n^3$\\
4-85 & $ee12|ee2|34|456|56|ee|ee|$ & 1/64                      &  315 & $1744 + 388n + 52n^2 + 3n^3$\\
4-86 & $ee12|ee3|ee3|44|56|e66|e|$ & 1/12                      &  315 & $1480 + 608n + 96n^2 + 3n^3$\\
4-87 & $ee12|ee3|ee3|45|ee6|666||$ & 11/288 -- log[4/3]/8      &  210 & $1272 + 816n + 96n^2 + 3n^3$\\
4-88 & $ee12|ee3|ee3|45|e56|66|e|$ & log[4/3]/6                & 1260 & $1592 + 556n + 38n^2 + n^3$\\
4-89 & $ee12|ee3|ee3|45|e66|e66||$ & 0.064928905166965         &  315 & $1488 + 608n + 90n^2 + n^3$\\
4-90 & $ee12|ee3|ee3|45|556|6|ee|$ & 1/16                      &  315 & $1480 + 608n + 96n^2 + 3n^3$\\
4-91 & $ee12|ee3|ee4|ee5|566|66||$ & 0.000297739052128         &  315 & $1280 + 816n + 90n^2 + n^3$\\
4-92 & $ee12|ee3|ee4|e45|e6|666||$ & ( -- 32321/540 -- 577 log[2] &  420 & $1284 + 816n + 87n^2$\\
&& + 17 log[3]/2 + 280 log[5])/81&&\\
4-93 & $ee12|ee3|ee4|e45|56|66|e|$ & 0.04307075173695          & 2520 & $1604 + 552n + 31n^2$\\
4-94 & $ee12|ee3|ee4|e45|66|e66||$ & 0.057786471957980         & 1260 & $1496 + 604n + 86n^2 + n^3$\\
4-95 & $ee12|ee3|ee4|e55|e66|66||$ & 0.104584627154817         &  315 & $1288 + 656n + 216n^2 + 27n^3$\\
4-96 & $ee12|ee3|ee4|e55|566|6|e|$ & 0.071658392755453         & 1260 & $1500 + 604n + 83n^2$\\
4-97 & $ee12|ee3|ee4|e55|666|e6||$ & 0.000132225484595         &  420 & $1284 + 816n + 87n^2$\\
4-98 & $ee12|ee3|ee4|e56|e56|66||$ & 0.053569617411952         &  630 & $1500 + 604n + 83n^2$\\
4-99 & $ee12|ee3|ee4|e56|556|6|e|$ & 0.042501325624530         & 2520 & $1604 + 552n + 31n^2$\\
4-100 & $ee12|ee3|ee4|445|6|e66|e|$ & 0.062614832018137        &  630 & $1596 + 556n + 35n^2$\\
4-101 & $ee12|ee3|ee4|455|56|6|ee|$ & 209/4608                 & 1260 & $1592 + 556n + 38n^2 + n^3$\\
4-102 & $ee12|ee3|ee4|455|66|e6|e|$ & 0.057786471957980        &  630 & $1600 + 552n + 34n^2 + n^3$\\
4-103 & $ee12|ee3|ee4|456|56|e6|e|$ & 0.020557711786           & 1260 & $1760 + 422n + 5n^2$\\
4-104 & $ee12|ee3|ee4|556|556||ee|$ &( -- 113 + 15$\pi^2$)/576  &  315 & $1488 + 608n + 90n^2 + n^3$\\
4-105 & $ee12|ee3|ee4|556|566|e|e|$ & 0.057602237765904        &  630 & $1604 + 552n + 31n^2$\\
4-106 & $ee12|ee3|e34|e4|56|e66|e|$ & 5/162                    & 1260 & $1660 + 492n + 35n^2$\\
4-107 & $ee12|ee3|e34|e5|ee6|666||$ & 4909/11664 + 61 log[2]/18 & 840 & $1380 + 768n + 39n^2$\\
&& + log[3]/12 -- 16 log[5]/9&&\\
4-108 & $ee12|ee3|e34|e5|e56|66|e|$ & 0.025133569519039        & 5040 & $1700 + 472n + 15n^2$\\
4-109 & $ee12|ee3|e34|e5|e66|e66||$ & 0.035640962304268        & 1260 & $1560 + 540n + 86n^2 + n^3$\\
4-110 & $ee12|ee3|e34|e5|556|6|ee|$ & 0.028534108723841        & 1260 & $1660 + 492n + 35n^2$\\
4-111 & $ee12|ee3|e34|45|e6|e66|e|$ & 0.018537650414849        & 5040 & $1764 + 408n + 15n^2$\\
4-112 & $ee12|ee3|e34|45|56|e6|ee|$ & 0.011176653628259        & 5040 & $1828 + 344n + 15n^2$\\
4-113 & $ee12|ee3|e34|45|66|ee6|e|$ & 0.018537650414849        & 2520 & $1724 + 428n + 35n^2$\\
4-114 & $ee12|ee3|e34|55|ee6|66|e|$ & 0.035640962304268        & 1260 & $1544 + 544n + 96n^2 + 3n^3$\\
4-115 & $ee12|ee3|e34|55|e56|6|ee|$ & -- 65/1152 + 7 log[4/3]/27 & 2520 & $1720 + 428n + 38n^2 + n^3$\\
4-116 & $ee12|ee3|e34|55|e66|e6|e|$ & 0.035640962304268        & 2520 & $1656 + 492n + 38n^2 + n^3$\\
4-117 & $ee12|ee3|e34|56|ee5|66|e|$ & 0.025133569519039        & 2520 & $1692 + 476n + 19n^2$\\
4-118 & $ee12|ee3|e34|56|e55|6|ee|$ & -- 65/1152 + 7 log[4/3]/27 & 2520 & $1728 + 424n + 34n^2 + n^3$\\
4-119 & $ee12|ee3|e34|56|e56|e6|e|$ & 0.008397599554132        & 5040 & $1888 + 294n + 5n^2$\\
4-120 & $ee12|ee3|e34|56|556|e|ee|$ & 0.017290671286081        & 2520 & $1764 + 408n + 15n^2$\\
4-121 & $ee12|ee3|e44|e45|6|e66|e|$ & 0.037085740379675        & 2520 & $1636 + 504n + 47n^2$\\
4-122 & $ee12|ee3|e44|e55|e6|66|e|$ & 0.104584627154817        & 1260 & $1432 + 620n + 126n^2 + 9n^3$\\
4-123 & $ee12|ee3|e44|e55|56|6|ee|$ & 0.039515809496933        &  630 & $1568 + 520n + 90n^2 + 9n^3$\\
4-124 & $ee12|ee3|e44|e56|e5|66|e|$ & 0.071658392755453        & 2520 & $1572 + 568n + 47n^2$\\
4-125 & $ee12|ee3|e44|e56|55|6|ee|$ & 0.039515809496933        & 1260 & $1496 + 556n + 126n^2 + 9n^3$\\
4-126 & $ee12|ee3|e44|e56|56|e6|e|$ & 0.018000831843930        & 2520 & $1712 + 430n + 45n^2$\\
4-127 & $ee12|ee3|e44|455|6|e6|ee|$ & 0.037085740379675        & 1260 & $1628 + 508n + 51n^2$\\
4-128 & $ee12|ee3|e44|456|e|e66|e|$ & 19/243                   & 1260 & $1564 + 572n + 51n^2$\\
4-129 & $ee12|ee3|e44|456|5|e6|ee|$ & 0.020141452226374        & 2520 & $1700 + 440n + 47n^2$\\
4-130 & $ee12|ee3|e44|555|e6|6|ee|$ & 0.000132225484595        &  420 & $1284 + 816n + 87n^2$\\
4-131 & $ee12|ee3|e44|556|e5|6|ee|$ & 0.075942365596715        & 1260 & $1564 + 572n + 51n^2$\\
4-132 & $ee12|ee3|e44|556|e6|e6|e|$ & 0.071658392755453        & 2520 & $1572 + 568n + 47n^2$\\
4-133 & $ee12|ee3|e44|556|56|e|ee|$ & 0.036300814059535        & 1260 & $1636 + 504n + 47n^2$\\
4-134 & $ee12|ee3|e45|e45|e6|66|e|$ & 0.0152659449             & 2520 & $1776 + 398n + 13n^2$\\
4-135 & $ee12|ee3|e45|e45|56|6|ee|$ & 0.011309549540615        & 1260 & $1808 + 350n + 29n^2$\\
4-136 & $ee12|ee3|e45|e46|ee|666||$ & 0.000233450823695        &  420& $1380 + 768n + 39n^2$\\
4-137 & $ee12|ee3|e45|e46|e5|66|e|$ & 0.015992626485276        & 5040 & $1744 + 414n + 29n^2$\\
4-138 & $ee12|ee3|e45|e46|55|6|ee|$ & 0.025664756185216        & 2520 & $1668 + 488n + 31n^2$\\
4-139 & $ee12|ee3|e45|e46|56|e6|e|$ & 0.009598391473892        & 5040 & $1850 + 327n + 10n^2$\\
4-140 & $ee12|ee3|e45|445|6|e6|ee|$ & 0.016854227325864        & 2520 & $1764 + 408n + 15n^2$\\
4-141 & $ee12|ee3|e45|446|e|e66|e|$ & 0.034685049239056          & 2520 & $1668 + 488n + 31n^2$\\
4-142 & $ee12|ee3|e45|446|5|e6|ee|$ & 0.017669745324596          & 2520 & $1732 + 424n + 31n^2$\\
4-143 & $ee12|ee3|e45|446|6|ee6|e|$ & 0.024459585212457          & 2520 & $1692 + 476n + 19n^2$\\
4-144 & $ee12|ee3|e45|456|ee|66|e|$ & 0.024251335395708          & 2520 & $1700 + 472n + 15n^2$\\
4-145 & $ee12|ee3|e45|456|e5|6|ee|$ & 0.010445183225406          & 5040 & $1840 + 334n + 13n^2$\\
4-146 & $ee12|ee3|e45|456|e6|e6|e|$ & 0.00803416337              & 5040 & $1898 + 287n + 2n^2$\\
4-147 & $ee12|ee3|e45|466|ee|e66||$ & 121/576 + 17$\pi^2$/648    & 1260 & $1560 + 540n + 86n^2 + n^3$\\
&& + 38 $Li_2[-1/3]/27$&&\\
4-148 & $ee12|ee3|e45|466|e5|e6|e|$ & 0.015992626485276          & 5040 & $1776 + 398n + 13n^2$\\
4-149 & $ee12|ee3|e45|466|e6|ee6||$ & 0.023925239638591          & 2520 & $1700 + 472n + 15n^2$\\
4-150 & $ee12|ee3|e45|466|56|ee|e|$ & 0.027328814083503          & 2520 & $1668 + 488n + 31n^2$\\
4-151 & $ee12|ee3|e45|666|ee5|e6||$ & 0.000087219840724          &  840 & $1380 + 768n + 39n^2$\\
4-152 & $ee12|ee3|334|5|ee6|e66|e|$ & 0.045781475051677          & 1260 & $1636 + 512n + 39n^2$\\
4-153 & $ee12|ee3|334|5|e56|e6|ee|$ & 0.021734049853977          & 1260 & $1732 + 432n + 23n^2$\\
4-154 & $ee12|ee3|344|45|6|ee6|ee|$ & 47/1536                    &  630 & $1664 + 496n + 26n^2 + n^3$\\
4-155 & $ee12|ee3|344|55|e6|e6|ee|$ & 0.035640962304268          &  630 & $1640 + 496n + 48n^2 + 3n^3$\\
4-156 & $ee12|ee3|344|56|e5|e6|ee|$ & -- 65/1152 + 7 log[4/3]/27 & 2520 & $1752 + 412n + 22n^2 + n^3$\\
4-157 & $ee12|ee3|344|56|56|ee|ee|$ & 7/384                      &  630 & $1744 + 416n + 26n^2 + n^3$\\
4-158 & $ee12|ee3|345|45|e6|e6|ee|$ & 0.009436905494             & 1260 & $1868 + 308n + 11n^2$\\
4-159 & $ee12|ee3|345|46|ee|e66|e|$ & 0.019629207404974          & 1260 & $1756 + 412n + 19n^2$\\
4-160 & $ee12|ee3|345|46|e5|e6|ee|$ & 0.009209696800186          & 5040 & $1876 + 304n + 7n^2$\\
4-161 & $ee12|ee3|345|46|56|ee|ee|$ & 3/256                      & 1260 & $1816 + 348n + 22n^2 + n^3$\\
4-162 & $ee12|ee3|445|446||ee6|ee|$ & ( -- 145 + 18$\pi^2$)/768  &  157.5 & $1536 + 560n + 90n^2 + n^3$\\
4-163 & $ee12|ee3|445|456|e|e6|ee|$ & -- 7/360 + 14 log[2]/27    & 2520   & $1764 + 408n + 15n^2$\\
&& -- 161 log[3]/864 -- 7 log[5]/96&&\\
4-164 & $ee12|ee3|445|456|6|ee|ee|$ & 29/1536                    & 1260 & $1720 + 428n + 38n^2 + n^3$\\
4-165 & $ee12|ee3|445|466|e|ee6|e|$ & 0.034685049239056          & 1260 & $1660 + 492n + 35n^2$\\
4-166 & $ee12|ee3|445|566|e6|ee|e|$ & 0.034058739909935          &  630 & $1664 + 488n + 34n^2 + n^3$\\
4-167 & $ee12|ee3|456|456|ee|e6|e|$ & 0.008764416908763          & 1260 & $1888 + 294n + 5n^2$\\
4-168 & $ee12|e23|e3|45|ee6|e66|e|$ & 2/81                       & 2520 & $1716 + 440n + 31n^2$\\
4-169 & $ee12|e23|e3|45|e56|e6|ee|$ & 1/81                       & 1260 & $1804 + 356n + 27n^2$\\
4-170 & $ee12|e23|e4|ee4|56|e66|e|$ & 5/162                      & 2520 & $1676 + 484n + 27n^2$\\
4-171 & $ee12|e23|e4|ee5|ee6|666||$ & 0.000087219840724          &  420 & $1380 + 768n + 39n^2$\\
4-172 & $ee12|e23|e4|ee5|e56|66|e|$ & 0.027328814083503          & 5040 & $1692 + 476n + 19n^2$\\
4-173 & $ee12|e23|e4|ee5|e66|e66||$ & 0.039515809496933          & 2520 & $1544 + 544n + 96n^2 + 3n^3$\\
4-174 & $ee12|e23|e4|ee5|556|6|ee|$ & 0.029586081676179          & 2520 & $1676 + 484n + 27n^2$\\
4-175 & $ee12|e23|e4|ee5|566|e6|e|$ & 0.027328814083503          & 5040 & $1692 + 476n + 19n^2$\\
4-176 & $ee12|e23|e4|e45|e6|e66|e|$ & 0.015965685467107          & 5040 & $1780 + 392n + 15n^2$\\
4-177 & $ee12|e23|e4|e45|56|e6|ee|$ & 0.008957256127767         & 10080 & $1868 + 308n + 11n^2$\\
4-178 & $ee12|e23|e4|e45|66|ee6|e|$ & 0.015965685467107          & 5040 & $1764 + 400n + 23n^2$\\
4-179 & $ee12|e23|e4|e55|e56|6|ee|$ & 11/108 -- 8 log[4/3]/27    & 5040 & $1736 + 412n + 38n^2 + n^3$\\
4-180 & $ee12|e23|e4|e55|e66|e6|e|$ & 0.039515809496933          & 2520 & $1648 + 496n + 42n^2 + n^3$\\
4-181 & $ee12|e23|e4|e56|e56|e6|e|$ & 0.007617697226747          & 5040 & $1892 + 288n + 7n^2$\\
4-182 & $ee12|e23|e4|e56|556|e|ee|$ & 0.015443212053930          & 5040 & $1780 + 392n + 15n^2$\\
4-183 & $ee12|e23|e4|456|56|ee|ee|$ & 7/720                      & 1260 & $1852 + 316n + 19n^2$\\
4-184 & $ee12|e23|34|e5|ee6|e66|e|$ & 0.020141452226374          & 5040 & $1748 + 416n + 23n^2$\\
4-185 & $ee12|e23|34|e5|e66|ee6|e|$ & 0.020141452226374          & 5040 & $1748 + 416n + 23n^2$\\
4-186 & $ee12|e23|34|56|ee5|e6|ee|$ & 0.008957256127767          & 5040 & $1852 + 316n + 19n^2$\\
4-187 & $ee12|e23|34|56|e56|ee|ee|$ & 0.008011413670392          & 2520 & $1884 + 292n + 11n^2$\\
4-188 & $ee12|e23|44|ee5|e6|e66|e|$ & 0.039515809496933          & 2520 & $1616 + 508n + 60n^2 + 3n^3$\\
4-189 & $ee12|e23|44|ee5|56|e6|ee|$ & 11/108 -- 8 log[4/3]/27    & 1260 & $1696 + 428n + 60n^2 + 3n^3$\\
4-190 & $ee12|e23|44|e55|e6|e6|ee|$ & 0.039515809496933          & 2520 & $1616 + 508n + 60n^2 + 3n^3$\\
4-191 & $ee12|e23|44|e56|56|ee|ee|$ & -- 41/576 + 8 log[4/3]/27  & 1260 & $1752 + 396n + 38n^2 + n^3$\\
4-192 & $ee12|e23|45|ee4|e6|e66|e|$ & 0.015965685467107          & 5040 & $1772 + 396n + 19n^2$\\
4-193 & $ee12|e23|45|ee4|66|ee6|e|$ & 0.015965685467107          & 2520 & $1756 + 404n + 27n^2$\\
4-194 & $ee12|e23|45|ee6|ee5|66|e|$ & 11/108 -- 8 log[4/3]/27    & 2520 & $1720 + 416n + 48n^2 + 3n^3$\\
4-195 & $ee12|e23|45|ee6|ee6|e66||$ & (-- 3 + 16 log[2]          & 2520 & $1772 + 396n + 19n^2$\\
&& -- 11 log[3] + 3 log[5])/54&&\\
4-196 & $ee12|e23|45|ee6|e56|e6|e|$ & 0.007617697226747          & 5040 & $1876 + 296n + 15n^2$\\
4-197 & $ee12|e23|45|e45|e6|e6|ee|$ & 0.005830695492014          & 5040 & $1940 + 240n + 7n^2$\\
4-198 & $ee12|e23|45|e46|ee|e66|e|$ & 0.014318930502268          & 5040 & $1796 + 376n + 15n^2$\\
4-199 & $ee12|e23|45|e46|e5|e6|ee|$ & 0.006284605628282         & 10080 & $1916 + 260n + 11n^2$\\
4-200 & $ee12|e23|45|e46|e6|ee6|e|$ & 0.006284605628282         & 10080 & $1924 + 256n + 7n^2$\\
4-201 & $ee12|e23|45|e46|56|ee|ee|$ & 7/864                      & 2520 & $1864 + 300n + 22n^2 + n^3$\\
4-202 & $ee12|e23|45|e66|ee5|e6|e|$ & 11/108 -- 8 log[4/3]/27    & 2520 & $1760 + 400n + 26n^2 + n^3$\\
4-203 & $ee12|e23|45|e66|ee6|ee6||$ & -- 41/576 + 8 log[4/3]/27  & 1260 & $1784 + 380n + 22n^2 + n^3$\\
4-204 & $ee12|e23|45|456|ee|e6|ee|$ & 1/135                      & 2520 & $1900 + 276n + 11n^2$\\
4-205 & $ee12|e23|45|466|ee|ee6|e|$ & 0.014318930502268          & 2520 & $1780 + 384n + 23n^2$\\
4-206 & $ee12|e33|e34|5|ee6|e66|e|$ & 0.037085740379675          & 2520 & $1652 + 496n + 39n^2$\\
4-207 & $ee12|e33|e34|5|e66|ee6|e|$ & 0.037085740379675          & 2520 & $1652 + 496n + 39n^2$\\
4-208 & $ee12|e33|e44|e5|e6|e66|e|$ & 0.104584627154817          & 1260 & $1528 + 584n + 72n^2 + 3n^3$\\
4-209 & $ee12|e33|e44|e5|66|ee6|e|$ & 0.104584627154817          &  630 & $1432 + 620n + 126n^2 + 9n^3$\\
4-210 & $ee12|e33|e44|56|56|ee|ee|$ & 0.030132446707805          &  315 & $1656 + 480n + 48n^2 + 3n^3$\\
4-211 & $ee12|e33|e45|e4|e6|e66|e|$ & 0.037085740379675          & 5040 & $1660 + 492n + 35n^2$\\
4-212 & $ee12|e33|e45|e6|ee6|e66||$ & 0.036300814059535          & 2520 & $1660 + 492n + 35n^2$\\
4-213 & $ee12|e33|e45|e6|e55|6|ee|$ & 0.071658392755453          & 2520 & $1572 + 568n + 47n^2$\\
4-214 & $ee12|e33|e45|e6|e56|e6|e|$ & 0.018000831843930          & 5040 & $1760 + 406n + 21n^2$\\
4-215 & $ee12|e33|e45|45|e6|e6|ee|$ & 0.011322590002467          & 2520 & $1808 + 358n + 21n^2$\\
4-216 & $ee12|e33|e45|46|ee|e66|e|$ & 0.029958025282025          & 2520 & $1676 + 476n + 35n^2$\\
4-217 & $ee12|e33|e45|46|e5|e6|ee|$ & 0.012754884477300          & 5040 & $1792 + 374n + 21n^2$\\
4-218 & $ee12|e33|e45|46|e6|ee6|e|$ & 0.012754884477300          & 5040 & $1768 + 386n + 33n^2$\\
4-219 & $ee12|e33|e45|66|ee6|ee6||$ & 0.030132446707805          &  630 & $1560 + 528n + 96n^2 + 3n^3$\\
4-220 & $ee12|e33|345|e|ee6|e66|e|$ & 4/81                       & 1260 & $1628 + 516n + 43n^2$\\
4-221 & $ee12|e33|345|6|ee6|ee6|e|$ & 0.013842067352976          & 1260 & $1780 + 384n + 23n^2$\\
4-222 & $ee12|e33|445|e6|e6|ee6|e|$ & 0.036300814059535          & 2520 & $1660 + 492n + 35n^2$\\
4-223 & $ee12|e33|445|e6|56|ee|ee|$ & 0.075942365596715          &  630 & $1564 + 572n + 51n^2$\\
4-224 & $ee12|e33|445|56|e6|ee|ee|$ & 0.029958025282025          &  630 & $1668 + 480n + 39n^2$\\
4-225 & $ee12|e33|456|e4|ee|e66|e|$ & 19/243                     &  630 & $1564 + 572n + 51n^2$\\
4-226 & $ee12|e33|456|e4|56|ee|ee|$ & 5/288 + log [5/3]/128      & 1260 & $1740 + 420n + 27n^2$\\
4-227 & $ee12|e34|e34|ee|56|e66|e|$ & 2/81                       &  630 & $1700 + 464n + 23n^2$\\
4-228 & $ee12|e34|e34|e5|e6|e66|e|$ & 0.011322590002467          & 2520 & $1832 + 346n + 9n^2$\\
4-229 & $ee12|e34|e34|56|56|ee|ee|$ & 0.005882277895645          &  630 & $1948 + 228n + 11n^2$\\
4-230 & $ee12|e34|e35|ee|ee6|666||$ & 947/1458 + (148 log[2]     &  420 & $1404 + 756n + 27n^2$\\
&& + 2 log[3] -- 76 log[5])/27&&\\
4-231 & $ee12|e34|e35|ee|e56|66|e|$ & 0.021050665320974          & 2520 & $1724 + 452n + 11n^2$\\
4-232 & $ee12|e34|e35|ee|e66|e66||$ & 0.030132446707805          &  630 & $1576 + 524n + 86n^2 + n^3$\\
4-233 & $ee12|e34|e35|ee|556|6|ee|$ & 0.023322339961950          &  630 & $1700 + 464n + 23n^2$\\
4-234 & $ee12|e34|e35|e4|e6|e66|e|$ & 0.012754884477300         & 10080 & $1808 + 366n + 13n^2$\\
4-235 & $ee12|e34|e35|e6|ee5|66|e|$ & 0.011756097196969          & 5040 & $1816 + 362n + 9n^2$\\
4-236 & $ee12|e34|e35|e6|ee6|e66||$ & 0.011888013318995          & 5040 & $1808 + 366n + 13n^2$\\
4-237 & $ee12|e34|e35|e6|e55|6|ee|$ & 0.010929428448857          & 5040 & $1824 + 350n + 13n^2$\\
4-238 & $ee12|e34|e35|e6|e56|e6|e|$ & 0.004660131730330         & 10080 & $1976 + 210n + n^2$\\
4-239 & $ee12|e34|e35|45|e6|e6|ee|$ & 0.006349405102246          & 2520 & $1914 + 263n + 10n^2$\\
4-240 & $ee12|e34|e35|46|e5|e6|ee|$ & 0.005585592314094         & 10080 & $1950 + 233n + 4n^2$\\
4-241 & $ee12|e34|e35|46|e6|ee6|e|$ & 0.006847511971597          & 5040 & $1906 + 275n + 6n^2$\\
4-242 & $ee12|e34|e35|46|56|ee|ee|$ & 0.007643502650282          & 5040 & $1888 + 286n + 13n^2$\\
4-243 & $ee12|e34|e35|66|ee5|e6|e|$ & 0.011756097196969          & 5040 & $1808 + 366n + 13n^2$\\
4-244 & $ee12|e34|e35|66|ee6|ee6||$ & 0.014486233305810          & 1260 & $1772 + 396n + 19n^2$\\
4-245 & $ee12|e34|e56|ee5|ee6|66||$ & 0.0204776278               & 1260 & $1716 + 456n + 15n^2$\\
4-246 & $ee12|e34|e56|ee5|e56|6|e|$ & 0.006513602711906         & 10080 & $1918 + 265n + 4n^2$\\
4-247 & $ee12|e34|e56|ee5|e66|e6||$ & 0.012333968941018          & 2520 & $1776 + 382n + 29n^2$\\
4-248 & $ee12|e34|e56|ee5|556||ee|$ & 0.014322122305123          & 1260 & $1780 + 392n + 15n^2$\\
4-249 & $ee12|e34|e56|e45|e6|e6|e|$ & 0.005299982325608          & 5040 & $1948 + 236n + 3n^2$\\
4-250 & $ee12|e34|e56|e55|e66|e|e|$ & 0.029958025282025          & 1260 & $1684 + 472n + 31n^2$\\
4-251 & $ee12|e34|e56|e56|e56|e|e|$ & 0.004081559                & 2520 & $2006 + 181n$\\
4-252 & $ee12|e34|335|6|e56|ee|ee|$ & 0.013026377750089          & 2520 & $1796 + 376n + 15n^2$\\
4-253 & $ee12|e34|345|ee|e6|e66|e|$ & 0.013842067352976          & 2520 & $1804 + 372n + 11n^2$\\
4-254 & $ee12|e34|345|e6|e6|ee6|e|$ & 0.004971866553244          & 2520 & $1978 + 207n + 2n^2$\\
4-255 & $ee12|e34|345|e6|56|ee|ee|$ & 0.006413361170045          & 2520 & $1928 + 250n + 9n^2$\\
4-256 & $ee12|e34|355|ee|e66|e6|e|$ & 0.030132446707805          & 1260 & $1672 + 476n + 38n^2 + n^3$\\
4-257 & $ee12|e34|356|ee|ee5|66|e|$ & 0.021050665320974          & 1260 & $1716 + 456n + 15n^2$\\
4-258 & $ee12|e34|356|ee|e55|6|ee|$ & -- 41/576 + 8 log[4/3]/27  & 1260 & $1760 + 392n + 34n^2 + n^3$\\
4-259 & $ee12|e34|356|ee|e56|e6|e|$ & 0.006484168615633          & 2520 & $1920 + 262n + 5n^2$\\
4-260 & $ee12|e34|356|ee|556|e|ee|$ & 0.013220795118932          & 1260 & $1804 + 372n + 11n^2$\\
4-261 & $ee12|e34|356|e4|56|ee|ee|$ & 0.006108725710044          & 2520 & $1940 + 240n + 7n^2$\\
4-262 & $ee12|e34|356|e5|ee5|6|ee|$ & 0.007466300196058          & 2520 & $1896 + 282n + 9n^2$\\
4-263 & $ee12|e34|356|e5|e56|e|ee|$ & 0.005298588893334          & 5040 & $1962 + 223n + 2n^2$\\
4-264 & $ee12|e34|356|45|e6|ee|ee|$ & 0.006922357564356          & 2520 & $1904 + 270n + 13n^2$\\
4-265 & $ee12|e34|556|e45|e6|e|ee|$ & 0.012333968941018          & 2520 & $1808 + 366n + 13n^2$\\
4-266 & $ee12|234|56|ee4|56|ee|ee|$ & 1/128                      &  630 & $1880 + 284n + 22n^2 + n^3$\\
4-267 & $ee12|334|456|56|ee|ee|ee|$ & 5/384                     & 157.5 & $1792 + 368n + 26n^2 + n^3$\\
4-268 & $ee12|345|346|ee|ee|e66|e|$ & 0.013555048885969          &  210 & $1812 + 360n + 15n^2$\\
\end{longtable}
\end{center}

\newpage
\begin{center}
\renewcommand*{\arraystretch}{1.25}
\begin{longtable}{|l|l|l|l|l|}
\caption[Feynman graphs for effective coupling $g_{10}$]{Feynman graphs for effective coupling 
$g_{10}$} 
\label{tab-a2} \\
\hline \multicolumn{1}{|c}{\textbf{No.}} &
\multicolumn{1}{|c}{\textbf{Graph description,}} &
\multicolumn{1}{|c}{\textbf{Integral value $\times$ $(8\pi)^L$,}}
&\multicolumn{1}{|c}{\textbf{Symmetry}} &
\multicolumn{1}{|c|}{\textbf{Tensor factor $\times$ $3^{L + 4}$,}} \\
\multicolumn{1}{|c}{\textbf{}} & \multicolumn{1}{|c}{\textbf{Nickel's
notations}} &\multicolumn{1}{|c}{\textbf{$L$ -- number of loops}}
&\multicolumn{1}{|c}{\textbf{factor}} &\multicolumn{1}{|c|}{\textbf{$L$ --
number of loops}} \\ \hline
\endfirsthead

\multicolumn{5}{c}%
{\tablename\ \thetable{} Feynman graphs for effective coupling $g_{10}$} \\
\hline \multicolumn{1}{|c}{\textbf{No.}} &
\multicolumn{1}{|c}{\textbf{Graph description,}} &
\multicolumn{1}{|c}{\textbf{Integral value $\times$ $(8\pi)^L$,}} &
\multicolumn{1}{|c}{\textbf{Symmetry}} &\multicolumn{1}{|c|}{\textbf{Tensor factor $\times$ $3^{L + 4}$,}} \\
\multicolumn{1}{|c}{\textbf{}} & \multicolumn{1}{|c}{\textbf{Nickel's
notations}} & \multicolumn{1}{|c}{\textbf{$L$ -- number of loops}} &
\multicolumn{1}{|c}{\textbf{factor}} &\multicolumn{1}{|c|}{\textbf{$L$ --
number of loops}} \\ \hline
\endhead

\hline \multicolumn{5}{|r|}{{Continued on next page}} \\ \hline
\endfoot
\hline \hline
\endlastfoot
1-1 & $ee12|ee3|ee4|ee4|ee|$ & 5/64                                            &  11340 & $242 + n$\\
 &         &               &                                                                      & \\
2-1 & $ee11|23|ee4|ee5|ee5|ee|$ & 5/64                                         &  28350 & $644 + 84 n + n^2$\\
2-2 & $ee12|ee2|34|ee5|ee5|ee|$ & 1/32                                         &  28350 & $692 + 36 n + n^2$\\
2-3 & $ee12|ee3|ee4|ee5|e55|e|$ & 97/1296                                      &  56700 & $646 + 83 n$\\
2-4 & $ee12|ee3|ee4|e45|e5|ee|$ & 17/648                                       & 113400 & $698 + 31 n$\\
2-5 & $ee12|ee3|e34|e5|ee5|ee|$ & 1/48                                         &  56700 & $710 + 19 n$\\
2-6 & $ee12|ee3|e45|e45|ee|ee|$ & 1/54                                         &  56700 & $714 + 15 n$\\
&               &                &                                                                  & \\
3-1 & $ee11|22|34|ee5|ee6|ee6|ee|$ & 5/64                                      &  14175 & $1608 + 492n + 86n^2 + n^3$\\
3-2 & $ee11|23|ee3|45|ee6|ee6|ee|$ & 1/32                                      &  28350& $1808 + 344n + 34n^2 + n^3$\\
3-3 & $ee11|23|ee4|ee4|56|ee6|ee|$ & 1/32                                      &  14175 & $1832 + 332n + 22n^2 + n^3$\\
3-4 & $ee11|23|ee4|ee5|ee5|66|ee|$ & 5/64                                      &  14175 & $1712 + 440n + 34n^2 + n^3$\\
3-5 & $ee11|23|ee4|ee5|ee6|e66|e|$ & 97/1296                                   &  56700 & $1720 + 438n + 29n^2$\\
3-6 & $ee11|23|ee4|ee5|e56|e6|ee|$ & 17/648                                    &  56700 & $1856 + 318n + 13n^2$\\
3-7 & $ee11|23|ee4|e45|e6|ee6|ee|$ & 1/48                                      & 113400 & $1880 + 294n + 13n^2$\\
3-8 & $ee11|23|ee4|e55|ee6|e6|ee|$ & 97/1296                                   &  56700 & $1720 + 438n + 29n^2$\\
3-9 & $ee11|23|ee4|e56|ee5|e6|ee|$ & 17/648                                    & 113400 & $1856 + 318n + 13n^2$\\
3-10 & $ee11|23|ee4|e56|e56|ee|ee|$ & 1/54                                     &  56700 & $1896 + 282n + 9n^2$\\
3-11 & $ee11|23|ee4|456|ee|ee6|ee|$ & 1/32                                     &  28350 & $1832 + 332n + 22n^2 + n^3$\\
3-12 & $ee11|23|ee4|556|ee6|ee|ee|$ & 5/64                                     &  14175 & $1712 + 440n + 34n^2 + n^3$\\
3-13 & $ee11|23|e34|e5|ee6|ee6|ee|$ & 17/648                                   &  56700 & $1824 + 334n + 29n^2$\\
3-14 & $ee11|23|e45|e46|ee|ee6|ee|$ & 1/54                                     &  56700 & $1888 + 286n + 13n^2$\\
3-15 & $ee12|ee2|33|45|ee6|ee6|ee|$ & 1/32                                     &  14175 & $1800 + 348n + 38n^2 + n^3$\\
3-16 & $ee12|ee2|34|ee4|56|ee6|ee|$ & 1/64                                     &  14175 & $1928 + 236n + 22n^2 + n^3$\\
3-17 & $ee12|ee2|34|ee5|ee6|e66|e|$ & 19/648                                   &  28350 & $1852 + 324n + 11n^2$\\
3-18 & $ee12|ee2|34|ee5|e56|e6|ee|$ & 5/432                                    & 113400 & $1988 + 192n + 7n^2$\\
3-19 & $ee12|ee2|34|ee5|e66|ee6|e|$ & 19/648                                   &  56700 & $1852 + 324n + 11n^2$\\
3-20 & $ee12|ee2|34|e45|e6|ee6|ee|$ & 5/432                                    &  56700 & $1980 + 196n + 11n^2$\\
3-21 & $ee12|ee2|34|e56|e56|ee|ee|$ & 1/108                                    &  28350 & $2020 + 160n + 7n^2$\\
3-22 & $ee12|ee3|ee3|45|ee6|e66|e|$ & 1/36                                     &  56700 & $1864 + 318n + 5n^2$\\
3-23 & $ee12|ee3|ee3|45|e56|e6|ee|$ & 1/72                                     &  56700 & $1968 + 214n + 5n^2$\\
3-24 & $ee12|ee3|ee4|ee4|56|e66|e|$ & 5/96                                     &  28350 & $1772 + 412n + 3n^2$\\
3-25 & $ee12|ee3|ee4|ee5|ee6|666||$ & 11609/184320 -- 7 log[4/3]/32            &   9450 & $1452 + 732n + 3n^2$\\
3-26 & $ee12|ee3|ee4|ee5|e56|66|e|$ & -- 1015/13824 + 551 log[4/3]/1296        & 113400 & $1776 + 410n + n^2$\\
3-27 & $ee12|ee3|ee4|ee5|e66|e66||$ & 0.071839533099720                        &  28350 & $1616 + 490n + 81n^2$\\
3-28 & $ee12|ee3|ee4|ee5|556|6|ee|$ & 209/4096                                 &  28350 & $1772 + 412n + 3n^2$\\
3-29 & $ee12|ee3|ee4|e45|e6|e66|e|$ & 1021/7776 -- 61 log[4/3]/162             & 113400 & $1880 + 306n + n^2$\\
3-30 & $ee12|ee3|ee4|e45|56|e6|ee|$ & 379/31104                                & 226800 & $1984 + 202n + n^2$\\
3-31 & $ee12|ee3|ee4|e45|66|ee6|e|$ & 1021/7776 -- 61 log[4/3]/162             & 113400 & $1876 + 308n + 3n^2$\\
3-32 & $ee12|ee3|ee4|e55|e56|6|ee|$ & 19/486 -- 4 log[4/3]/81                  & 113400 & $1828 + 332n + 27n^2$\\
3-33 & $ee12|ee3|ee4|e55|e66|e6|e|$ & 0.071839533099720                        &  56700 & $1724 + 436n + 27n^2$\\
3-34 & $ee12|ee3|ee4|e56|e56|e6|e|$ & 0.011272671247075                        & 113400 & $1990 + 197n$\\
3-35 & $ee12|ee3|ee4|e56|556|e|ee|$ & -- 517/3456 + 97 log[4/3]/162            & 113400 & $1880 + 306n + n^2$\\
3-36 & $ee12|ee3|ee4|445|6|ee6|ee|$ & 163/6144                                 &  28350 & $1868 + 316n + 3n^2$\\
3-37 & $ee12|ee3|ee4|456|56|ee|ee|$ & 13/1024                                  &  28350 & $1980 + 204n + 3n^2$\\
3-38 & $ee12|ee3|e34|e5|ee6|e66|e|$ & -- 11/324 + 5 log[4/3]/27                & 226800 & $1900 + 284n + 3n^2$\\
3-39 & $ee12|ee3|e34|e5|e56|e6|ee|$ & 7/864                                    & 226800 & $2040 + 146n + n^2$\\
3-40 & $ee12|ee3|e34|45|e6|ee6|ee|$ & 185/20736                                & 113400 & $2032 + 154n + n^2$\\
3-41 & $ee12|ee3|e34|55|ee6|e6|ee|$ & -- 11/324 + 5 log[4/3]/27                & 113400 & $1888 + 290n + 9n^2$\\
3-42 & $ee12|ee3|e34|56|ee5|e6|ee|$ & 7/864                                    & 226800 & $2036 + 148n + 3n^2$\\
3-43 & $ee12|ee3|e34|56|e56|ee|ee|$ & 71/10368                                 & 113400 & $2072 + 114n + n^2$\\
3-44 & $ee12|ee3|e44|e55|e6|e6|ee|$ & 0.071839533099720                        &  56700 & $1724 + 436n + 27n^2$\\
3-45 & $ee12|ee3|e44|e56|e5|e6|ee|$ & 19/486 -- 4 log[4/3]/81                  & 226800 & $1864 + 314n + 9n^2$\\
3-46 & $ee12|ee3|e44|e56|56|ee|ee|$ & -- 11/162 + 8 log[4/3]/27                &  56700 & $1896 + 282n + 9n^2$\\
3-47 & $ee12|ee3|e45|e45|e6|e6|ee|$ & 0.006047499233891                        & 113400 & $2086 + 101n$\\
3-48 & $ee12|ee3|e45|e46|ee|e66|e|$ & 541/2592 -- 2 log[4/3]/3                 & 113400 & $1912 + 274n + n^2$\\
3-49 & $ee12|ee3|e45|e46|e5|e6|ee|$ & 0.007043504786243                        & 453600 & $2054 + 133n$\\
3-50 & $ee12|ee3|e45|e46|56|ee|ee|$ & 1/108                                    & 113400 & $2000 + 178n + 9n^2$\\
3-51 & $ee12|ee3|e45|446|e|ee6|ee|$ & -- 133/768 + 2 log[4/3]/3                &  56700 & $1904 + 282n + n^2$\\
3-52 & $ee12|ee3|e45|456|ee|e6|ee|$ & 79/10368                                 & 226800 & $2048 + 138n + n^2$\\
3-53 & $ee12|ee3|e45|466|ee|ee6|e|$ & 541/2592 -- 2 log[4/3]/3                 & 113400 & $1908 + 276n + 3n^2$\\
3-54 & $ee12|ee3|345|46|ee|ee6|ee|$ & 19/2048                                  &  28350 & $2028 + 156n + 3n^2$\\
3-55 & $ee12|ee3|456|456|ee|ee|ee|$ & 7/1024                                   &   9450 & $2076 + 108n + 3n^2$\\
3-56 & $ee12|e23|e4|ee5|ee6|e66|e|$ & 541/2592 -- 2 log[4/3]/3                 & 113400 & $1864 + 314n + 9n^2$\\
3-57 & $ee12|e23|e4|ee5|e56|e6|ee|$ & 1/108                                    & 226800 & $2012 + 172n + 3n^2$\\
3-58 & $ee12|e23|e4|e56|e56|ee|ee|$ & 1/144                                    & 113400 & $2056 + 130n + n^2$\\
3-59 & $ee12|e23|45|ee6|ee6|ee6|e|$ & 1/144                                    & 113400 & $2052 + 132n + 3n^2$\\
3-60 & $ee12|e33|e45|e6|ee6|ee6|e|$ & -- 11/162 + 8 log[4/3]/27                & 113400 & $1908 + 276n + 3n^2$\\
3-61 & $ee12|e34|e35|e6|ee5|e6|ee|$ & 0.005150554023338                        & 226800 & $2102 + 85n$\\
3-62 & $ee12|e34|e35|e6|ee6|ee6|e|$ & 0.005381671263319                        &  75600 & $2094 + 93n$\\
3-63 & $ee12|e34|e56|ee5|ee6|e6|e|$ & 0.006027278683541                        &  75600 & $2070 + 117n$\\
3-64 & $ee12|e34|356|ee|e56|ee|ee|$ & 29/5184                                  &  28350 & $2096 + 90n + n^2$\\
\end{longtable}
\end{center}